\documentclass[a4paper,10pt,preprint,3p]{elsarticle}

\usepackage{lineno}
\modulolinenumbers[5]

\usepackage{listings}
\usepackage{color}
\usepackage{graphicx}
\usepackage{float}
\usepackage{graphicx}
\usepackage{natbib} 
\usepackage{url}
\usepackage[obeyFinal]{todonotes}
\usepackage{float}
\usepackage{listings}
\usepackage{rotating}
\usepackage{multirow}
\usepackage{multicol}
\usepackage{booktabs}
\usepackage{enumitem}
\usepackage{smartdiagram}
\usepackage{longtable}
\usepackage[parfill]{parskip} 
\usepackage{array,etoolbox}
\usepackage{tablefootnote}

\usepackage[tikz]{mdframed}
\definecolor{light-gray}{gray}{0.95}\newmdenv[%
outerlinewidth=2,%
roundcorner=10pt,%
linecolor=gray,%
backgroundcolor=light-gray,%
]{myframe}

\usepackage{tikz}
\usetikzlibrary{shapes,arrows,positioning}

\usepackage{xurl}
\usepackage[colorlinks=true,allcolors=black,citecolor=blue, linkcolor=blue]{hyperref}

\usepackage{helvet}

\preto\tabular{\setcounter{magicrownumbers}{0}}
\newcounter{magicrownumbers}

\usepackage{sectsty}
\sectionfont{\normalsize}
\subsectionfont{\normalsize}
\subsubsectionfont{\normalsize}
\paragraphfont{\normalsize}

\newcommand{\ie}{i.e.}
\newcommand{\eg}{e.g.}

\newcommand{\Reffig}[1]{Figure~\ref{#1}}

\newcommand{\Reftab}[1]{Table~\ref{#1}}

\definecolor{gray}{rgb}{0.4,0.4,0.4}
\definecolor{darkblue}{rgb}{0.0,0.0,0.6}
\definecolor{cyan}{rgb}{0.0,0.6,0.6}
\definecolor{mauve}{rgb}{0.58,0,0.82}

\lstdefinelanguage{XML}
{
	frame=tb,
	aboveskip=3mm,
	belowskip=3mm,
	breaklines=true,
	breakatwhitespace=true,
	morestring=[b]",
	morestring=[s]{>}{<},
	morecomment=[s]{<?}{?>},
	stringstyle=\color{darkblue},
	identifierstyle=\color{darkblue},
	keywordstyle=\color{mauve},
	morekeywords={xmlns,version,type,name,codeSpace}
}

\journal{Automation in Construction}

\begin{document}
	
\begin{frontmatter}
		
	\title{IFC models for (semi)automating common planning checks for building permits}
		
		\author[tud]{Francesca Noardo}
		\ead{f.noardo@tudelft.nl}
		\author[tud]{Teng Wu}
		\author[tud]{Ken Arroyo Ohori}
		\author[tud]{Thomas Krijnen}
		\author[tud]{Jantien Stoter}
		\address[tud]{3D Geoinformation, Delft University of Technology, Delft, The Netherlands}
		

\begin{myframe}
			
			This is the author's version of the work. 
			
			It is posted here only for personal use, not for redistribution and not for commercial use.
			
			The definitive version is published in the journal \emph{Automation in construction}.
			\\	
			\\
			Noardo, F., Wu, T., Arroyo Ohori, K., Krijnen, T., Stoter, J. (2022). IFC models for semi-automating common planning checks for building permits. \emph{Automation in construction}.
			\\\textsc{doi}: \url{https://doi.org/10.1016/j.autcon.2021.104097}
			\\
			
\end{myframe}

\begin{abstract}
To support building permit issuing with automatic digital tools, the reuse of models produced by designers would make the process quicker and more objective. However, current studies and pilots often leave a gap with respect to the models as actually provided by architects, having varying quality and content. In this study, rather than taking a top down approach, we started from the available data and made the necessary inferences, which gave the opportunity to tackle basic and common issues often preventing smooth automatic processing. Specific characteristics of the IFC models were outlined and a tool was developed to extract the necessary information from them to check representative regulations. While the case study is specific in location, regulations and input models, the type of issues encountered are a generally applicable example for automated code compliance checking. This represents a solid base for future works towards the automation of building permits issuing.
\end{abstract}

\begin{keyword}
	BIM, Building Information Modelling, Industry Foundation Classes, urban planning compliance, digital building permit, data quality
\end{keyword}

\end{frontmatter}

\section{Introduction}\label{sec:intro}

Digitalization of administration data and procedures is of current interest all over the world \cite{parviainen2017tackling}, in order to increase efficiency in bureaucratic procedures and reduce inaccuracies, irregularities and ambiguities, besides reducing the use of resources for more sustainable practices \citep{samasoni2014exploratory}.
Several pilots are implemented and currently used throughout the world (\eg\ in Finland\footnote{\url{https://kirahub.org/en/home/} and \url{https://aec-business.com/how-bim-is-revolutionizing-building-control-in-finland/} Accessed 3rd November 2021}, Norway\footnote{\url{https://www.ks.no/ebyggesak} Accessed 3rd November 2021}, Estonia\footnote{\url{https://aec-business.com/digital-transformation-of-the-estonian-construction-sector-an-interview-with-jaan-saar/} Accessed 3rd November 2021}, Germany\footnote{\url{https://www.bauministerkonferenz.de/verzeichnis.aspx?id=19967&o=759O19967} Accessed 3rd November 2021}, Slovenia\footnote{\url{https://www.projekt.e-prostor.gov.si} Accessed 3rd November 2021} and many more).
Within such an effort, the development towards digitalization and automation of building permit process---the procedure for approval of a building plan by a public authority---is widely promoted and studied in Europe and around the world \citep{Noardo19b,noardo2020integrating,eastman2009automatic,lee2016translating,beach2019d,guler2021reformative}.
Moreover, the European Directive 2014/24/EU\footnote{\url{https://eur-lex.europa.eu/legal-content/EN/TXT/?uri=celex\%3A32014L0024} Accessed 3rd November 2021} strongly encourages the use of BIM for public projects.
Especially big cities, such as Rotterdam in the Netherlands, report a high number of requested permits and often have to deal with complex situations that could benefit from a (semi-)automatic tool to at least solve the most straightforward checks.
This would allow the municipality officers to focus better on cases of non-straightforward compliance and possible exemptions, besides reducing the time and resources necessary for each building permit procedure and have advantage of more objective checking systems, not biased by human judgment and personal interpretations.

Digitalization implies passing from (analogue) document-based systems to digital data-driven processes.
Essential condition is therefore the use of 3D information systems as input, \ie\  mainly building information models (BIMs).
BIMs were developed in the architectural engineering and construction domain, to assist the building design, besides having features useful to project management and asset and facility management \citep{mcglinn2016identifying,wu2018building}.
The Industry Foundation Classes (IFC) by buildingSMART\footnote{\url{https://www.buildingsmart.org/standards/bsi-standards/industry-foundation-classes/} Accessed 3rd November 2021} are the well-known open standard used in most cases for exchanging BIMs.
Open formats are usually required by public administrations, besides being the preferable choice for the sake of interoperability and replicability.
Therefore, the IFC format was considered in this investigation.

\subsection{Related work}\label{sec:relwork}

Building permit process digitalization is part of the broader framework of the digitalization in the construction sector, starting with the need of providing digital information about the city context \citep{salheb2020automatic,isprs-archives-XLIV-4-W1-2020-41-2020} and the related planning regulations \citep{beach2018semantic,hjelseth2011capturing,moult2020compliance,marchant2016design,lee2016translating,park2015rule} and ending in the digital building logbook \citep{signorini2021digital} and digital asset and facility management tools \citep{dejaco2020building,cavka2017developing,al2021integrated,silva2021facility}.
Finally, the delivered models should be a suitable input to keep the city data up-to-date \citep{biljecki2021extending,eriksson2020requirements,guler2021reformative}.

Many examples and valuable studies exist about the use of BIM, to perform automatic compliance checks for building regulations \citep{choi2017development,clayton2013automated,solihin2020simplified,kim2016development,ciribini2016implementation}.
The use of IFC is central to many of them \citep{pauwels2011semantic,lee2015automated,kim2020development,temel2020investigation}
Moreover, other works are intended to use the integrated information coming both from the BIM (likely in IFC) and the geoinformation related to the context of the planned building, to support the automation of the checks \citep{olsson2018automation}.

\subsection{Gap in research and relevance of this work}

However, the topic is wide and complex: the many components and data involved often contain many sub-issues and uncertainties in their turn.
This has prevented many of the performed studies (some of them started more than 10 years ago, \eg\ \citep{Tan2010a,Nguyen2011}) from being implemented by municipalities in a real-world production environment.
In particular, the use of data coming from the current practice (as opposed to data specially made for this purpose) often still appears to be tricky \citep{Arroyo-Ohori18a,Berlo13,barbini2019bim}.

In order to address these challenges, we chose to adopt a bottom-up approach (Section~\ref{sec:methodology}), investigating the readiness of practice to support such a procedure, which is the most basic premise for enabling building permits digitalization, although being little investigated so far.
Many of the current efforts (Section \ref{sec:relwork}) focus on the technological developments, considering the IFC data model as defined by the standard.
The models used for testing are often modeled on purpose, within academic environment and caring their validity for the scope of the test.
However, this is quite far from how IFC models are produced in practice \citep{noardo2021inspection} and, as a consequence, the application of the output of many studies is hindered in practice.
Practitioners have often little control on the quality and content of IFC models. Because they sometimes lack specific knowledge and because they work with commercial systems that do the conversions for them, allowing little customization.

The main focus of this study was, therefore, the use of models produced by practitioners to design real buildings recently approved or waiting for approval.
Within a project carried out with the city of Rotterdam\footnote{\url{https://3d.bk.tudelft.nl/projects/rotterdamgeobim\_bp/} Accessed 3rd November 2021} we had the opportunity of investigating the overall use case and the several related issues in practice, \ie\ valid regulations, a real-world building permit checking case, existing municipal data, designs as submitted by architects, practice of permit-checkers, and so on.

We could inspect and analyze two delivered models by architects as a significant sample in order to check their compliance with the general IFC prescriptions (Section~\ref{sec:results1}).
The consequences of the possible misalignment on the development of an automatic tool to assist in urban planning checks were pointed out.
In order to ensure the validity of our results beyond the scope of the local project, we had made, in parallel, a wider inspection of IFC models provided by practitioners, from different countries and professionals, to ensure that similar characteristics are found in most of such models and those reported here are not exceptional \citep{noardo2021inspection}.

In addition, a demonstrator tool able to assist in checking some regulations was implemented, that could use the provided models (and therefore most of the current IFC models as they are usually exported by authoring software) notwithstanding their issues (Section~\ref{sec:restool}).
In particular, according to the findings about the state of the models, the tool used a mostly geometry-based approach, associated with the minimum semantics useful to select the interested parts of the model (\eg\ division in storeys, \textit{Elevation} attribute of \textit{IfcBuildingStorey} etc.).
According to our experience, guidelines to modelers are proposed in Section~\ref{sec:resguidelines}.

Although the findings explained in the paper are very practical, they represent a concrete starting point, investigated with a systematic and scientific method (described in Section \ref{sec:methodology}), that deserves to be explained and clarified.
It supports further research and application development about using IFC and automating and digitalizing the building permits checks.
Several research topics can be developed from the outcomes of this study, among which: the need of defining clear data requirements, as most useful for the use and analysis of the models; their formal representation within the IFC model, most likely by means of a specific Model View Definition; IFC validation; the selection and extraction of the needed information from the overall model, as well as the kind of encoding, in order to work with smaller data sets; the connection of BIM with geoinformation to improve automation and enhance analysis performance; methods to fix the not suitable models by inferring the needed information; methods to generate and extract the needed geometry by selecting and possibly generalizing the geometries of elements composing the BIM.

The issues investigated and addressed in the paper are sometimes part of the experience of professionals and researchers dealing with BIM from practice, but there was very little evidence of this in the scientific literature, bringing to the wrong belief that any IFC-compliant model, exported by any IFC-Certified software could be directly used within any other IFC-based tool.
Such aim is instead still far, although some work is in progress with respect to it.
For example, many of the institutions experimenting with digital building permits (see Section \ref{sec:intro}) are developing data requirements specifications for the IFC models to be delivered.

The overall topic of digitalization of building permits is vast and several sub-issues must be tackled before getting to a comprehensive solution.
The EUnet4DBP\footnote{\url{https://3d.bk.tudelft.nl/projects/eunet4dbp/about.html} Accessed 4th November 2021} defined the implied ambitions to aim for in order to enable digital building permit. Related requirements are also proposed to reach such ambitions.
They, together, represent a good overview on the thorough topic and summary of the related issues.
The work presented in this paper contributes to the ambitions defined as T3 -- Technologies for data visualization, data analysis and data manipulation -- and R2 -- Explicit specification of data requirements. Moreover, the requirement r11, related to the interoperability is addressed.
Those are currently among the most investigated issues in the overall topic, since automatic checking of regulation is seen as the core premise for digitalization of the whole process \citep{noardo2021unveiling}.
However, the approach proposed in this paper brings a different perspective and points out new evidences, essential to develop new critical progress on the topic.

\section{Methodology}\label{sec:methodology}

A bottom-up approach was chosen to investigate the readiness of practice and to identify remaining issues.
The definition of a methodology itself was formulated following iterative processes including consultations and feedback collection in a strong collaboration with municipality officers \citep{Noardo3DGeoInfo2020rotterdam}.
The steps of the overall methodology for the whole project are as follows:

\begin{enumerate}
	\item Selection of a case study (Section~\ref{sec:cs}) 
	and two regulations to point out the necessary detailed steps and issues to be investigated;
	\item Interpretation and formalization of two regulations\footnote{Investigating the Automation of the building permission issuing process through 3D GeoBIM information --- Deliverable 1 \url{https://3d.bk.tudelft.nl/projects/rotterdamgeobim_bp/Deliverable1_v2.pdf} Accessed 4th November 2021} \citep{Noardo3DGeoInfo2020rotterdam};
	\item Analysis of the available data produced by practice, to assess their suitability for checking the compliance to regulations by an automatic tool, and description of the most problematic aspects;
	\item Implementation and test of a demonstrator tool relying on the IFC models as they are (\ie\ coping with the weaknesses pointed out in the previous step) to check the regulations;
	\item Scaling up the experience: guidelines and recommendations from the lessons learnt.
\end{enumerate}

In this paper, the last three steps are addressed (Sections~\ref{sec:metifcpractice}, \ref{sec:mettool} and~\ref{sec:metglgeoref} respectively) with main reference to the case study described in Section \ref{sec:cs}.
We addressed the Step 2 in a separate publication \citep{Noardo3DGeoInfo2020rotterdam}.

\subsection{Using IFC models from practice}\label{sec:metifcpractice}

We started by inspecting and analyzing how the necessary information is stored in the models that were submitted to the municipality from two different architecture firms.

The practice of building design and BIM follows criteria and best practices defined both within the design discipline itself and sometimes also by national guidelines.
Meanwhile, the buildingSMART IFC\footnote{\url{https://technical.buildingsmart.org/standards/ifc/} Accessed 4th November 2021} data model is defined as the reference standard for OpenBIM, which is therefore the desirable interoperable format to be used by all tools.
However, designers can follow their own specific unofficial rules when modeling, and on the other hand, software can implement IFC in different ways (outside the influence of designers), so that we cannot straightforwardly expect the same exported product 
(\ie\ complying with exactly the same part of the IFC structure, used in the same way) every time and in every case.

Only considering the IFC schema, assumptions can be made on which to base the computation of the needed geometry.
For example, one can consider the modeling of storeys and their reference bounding boxes useful to calculate the building maximum extension, as well as to segment the building envelope.
Moreover, a BIM is usually composed of several discipline-separated IFC models (\ie\ the one storing architectural elements, the structural one, the installations and so on).
It is necessary to know what information is found in each of them in order to calculate the correct geometry and envelope of the building: for example, all the facade articulations or the internal building elements, to compute paths and volumes and internal subdivisions.
The measurement of some of such elements could be a reference when trying to infer the geometric features by tools using a simplified process.
Another case is the assumption that the building elements are consistently modeled, with semantics correctly associated, as prescribed by IFC\@.
However, those assumptions could be misleading and far from what is the design and modeling practice of architects (Section~\ref{sec:results1}).

The models were manually inspected, with several aims: to outline the actual characteristics of the BIMs, both as consequence of the modeling practice and of the implementation and use of the IFC standard; to point out the related possible issues for their use within automatic tools; and to potentially guide the implementation.
Commonly adopted BIM software and viewers were used for the inspection, namely \textit{Autodesk Revit}\footnote{\url{https://www.autodesk.com/products/revit/overview} Accessed 4th November 2021}, the \textit{Solibri Model Viewer}\footnote{\url{https://www.solibri.com/solibri-anywhere} Accessed 4th November 2021}, \textit{RDF IFC viewer}\footnote{\url{http://rdf.bg/product-list/ifc-engine/ifc-viewer/} Accessed 4th November 2021}.


\subsection{The checking tool}\label{sec:mettool}

Based on the results of the inspection, a semi-automatic tool to check regulations was implemented.
It was developed iteratively, based on the data typically present in IFC files, attempting to derive the required information based on a mixture of the geometry and semantics in the models.
The tool is open source and available at \url{https://github.com/tudelft3d/GEOBIM\_Tool}.

While the tool is currently suited to the specific regulations implemented and their particularities, it shows general methods that can be reused for the implementation of other regulations.
Currently, it supports checking of the dimension regulation, by means of the measurement of maximum height (Section~\ref{sec:maxh}); extraction of storeys profiles and calculation of reciprocal overlap (Section~\ref{sec:algovcalc}) 
and measurement of overhangs of the top part of the building with respect to the base (Section~\ref{sec:algovdist}).
The results of the processing tool were checked against the ground truth, as measured from the IFC models manually, by means of BIM and IFC viewers.

The implementation of an automated tool assisting with the parking regulation was currently deemed very difficult, for the reasons later explained.


\subsection{Guidelines}\label{sec:metglgeoref}

Several guidelines must be provided in order for modelers to provide suitable data for automatic tools.
The plain export of data to the IFC format is not sufficient: first, it is necessary to take care that information is not lost or damaged in the conversion from proprietary software formats \citep{noardo2021reference}.
Moreover, the choice of consistent semantics and a comprehensive filling of the information necessary for the use case is important, as well as the correct grouping of storeys and representation of spaces and apartments \citep{noardo2021inspection}.
The critical aspects highlighted during the inspection of the models and revealed to be relevant for the implementation are listed as guidelines, addressed mainly to building information modelers, with a few suggestions also for city planners and geoinformation providers.

\subsection{The case study and the chosen regulations}\label{sec:cs}

In consultation with the Rotterdam team, the planning zone \textit{``Centrum 3''} of the \emph{Waterstad bestemmingsplan} (destination plan) in the centre of Rotterdam was selected.
The dimension regulation\footnote{\url{https://www.ruimtelijkeplannen.nl/documents/NL.IMRO.0599.BP1054Waterstad-va01/r_NL.IMRO.0599.BP1054Waterstad-va01.html\#\_5\_Centrum-3} Accessed 4th November 2021} and the parking regulation\footnote{\url{http://decentrale.regelgeving.overheid.nl/cvdr/xhtmloutput/Historie/Rotterdam/486392/486392\_1.html}, Accessed 4th November 2021, and \url{https://www.ruimtelijkeplannen.nl/documents/NL.IMRO.0599.BP1054Waterstad-va01/b_NL.IMRO.0599.BP1054Waterstad-va01\_rb2.pdf} Accessed 4th November 2021} were considered for the project.
They were selected with the city-planners as most representative examples needing geometry processing and measurements very common in many city regulations (making the work worth to be replicated) and having advantage of the integration of BIM with geoinformation.

Solving the ambiguity points left by the dimension regulation text (here translated from Dutch) was the initial step of the project \citep{Noardo3DGeoInfo2020rotterdam}:
\textit{“The maximum building height is 100 meters, on the understanding that it can be realized with a lower building body of a maximum of 17 meters high and a construction, of a maximum of 50\% of the surface of such building body, on the top of it. At the location of Boompjes 60--68 and Boompjes 55--58, an overhang of 5 meters on the Boompjes side and 10 meters on the Hertekade side is permitted”.}

The interpretation of regulation regarding the provision of parking places depending on the area of living units, was easier, since the logic was already formalized by the municipality officer in a spreadsheet application.
Nevertheless, in real municipality practice, expert officers know that other factors (documented by further rules and laws or customs) have an influence on the count of parking places to be provided, adding further complexity.
To check the regulation it is necessary to:
(1) Define the building units and their function; (2) Calculate the area of each apartment; (3) Calculate the minimum number of the parking units; (4) Check if the planned parking units meet the regulation.

\begin{figure}[H]
	\centering
	\includegraphics[width=0.6\linewidth]{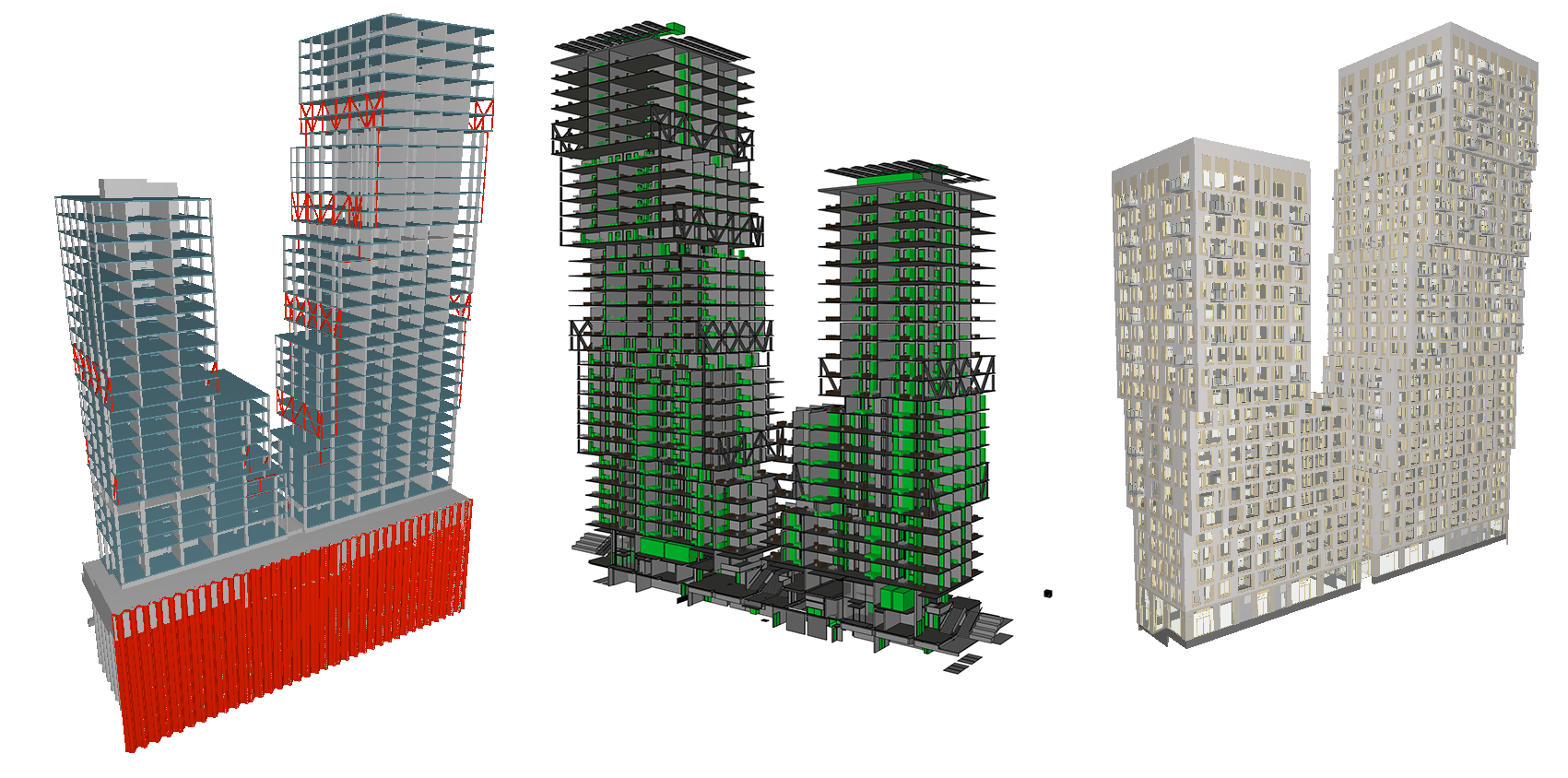}(a)
	\includegraphics[width=0.4\linewidth]{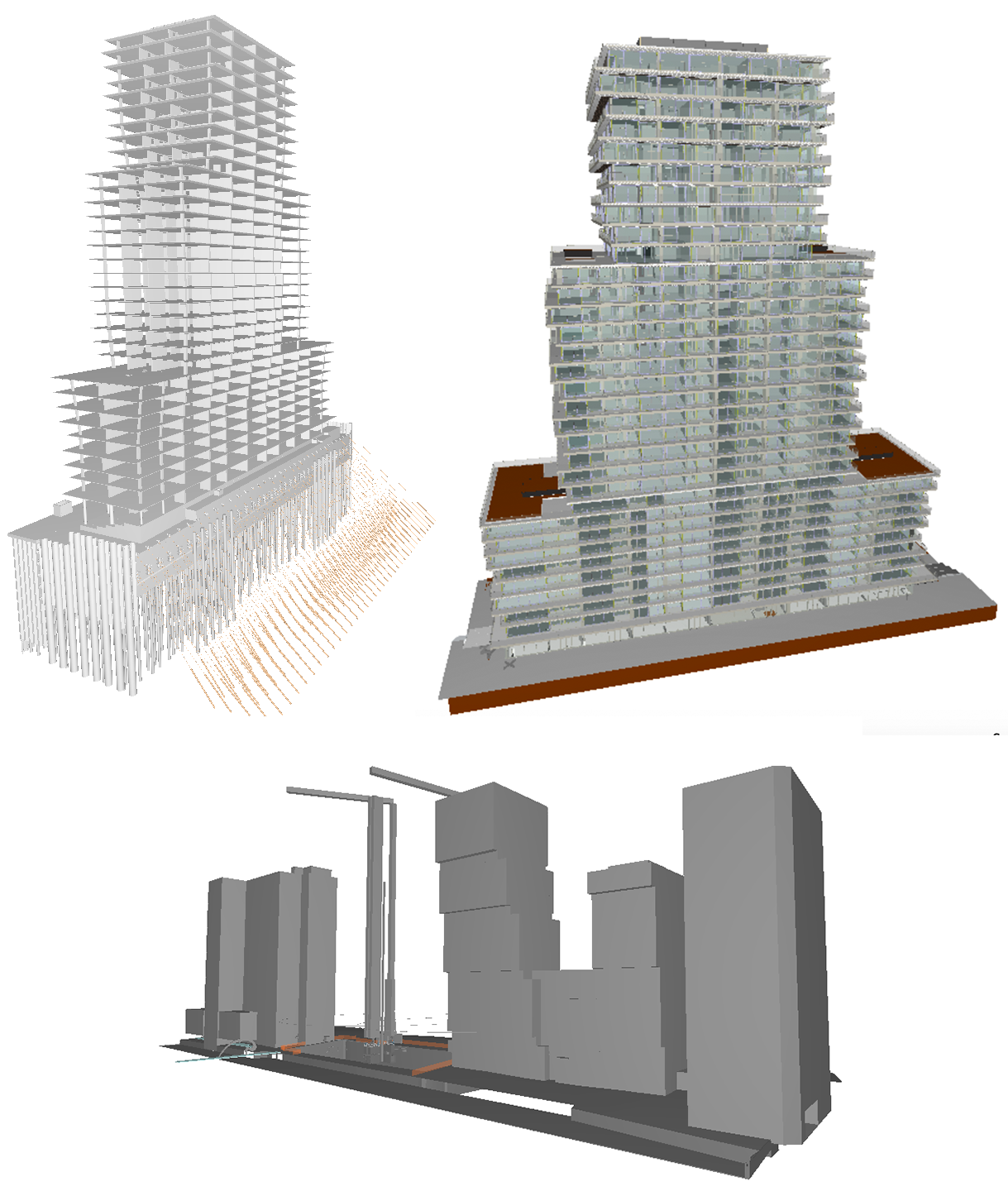}(b)
	\caption{(a) The three IFC models composing the BIM, from left to right, the structural, the architectural and the facades one. (b) The three IFC models composing the BIM, clockwise, the structural, the architectural and the context one.}%
	\label{fig:terraceBIM}
\end{figure}

Within the area, two buildings were recently designed, that were used as specific case studies for this project.
The two respective BIMs were kindly provided in IFC (v.2x3) for the tests.
The two models were delivered by the architects without following any requirements or Level of Information and geometry specification.
Although providing previously such recommendations could solve some of the issues detected in this study, those models were chosen because, at present, the models are usually not following any requirements specifications but are being delivered as automatically exported by the authoring software.
It could therefore help in formulating such Level of Information Need more effectively and aligned with practice.
The first is the \textit{Peak Tower}\footnote{Project organization: Peak Development --- Architect: Team V Architectuur}, composed of a structural, an architectural and a facades model (\Reffig{fig:terraceBIM}(a)).
They are correctly registered together, by means of the same reference point and orientation.
They are not georeferenced, though: the \textit{IfcCartesianPoint} referenced by \textit{IfcSite}
, reports coordinates \((0,0,0)\), whilst the \textit{RefLatitude} and \textit{RefLongitude} attributes in the \textit{IfcSite} would locate the model in a general location in Amsterdam.
The other BIM, representing the \textit{Terraced tower}\footnote{Project organization: Provast --- Architect: OZ Architects} (\Reffig{fig:terraceBIM}(b)), is composed of a structural and an architectural model, plus one representing the context and elements probably belonging to the work site.
They are correctly registered together too and the direction of the local Cartesian system of the model is correctly stored as \textit{TrueNorth} attribute of the \textit{IfcGeometricRepresentationContext}.
In this case the \textit{RefLatitude} and \textit{RefLongitude} attributes in the \textit{IfcSite}
would locate the model in the United States, but the \textit{IfcCartesianPoint} referenced by \textit{IfcSite} reports the right coordinates of the Rotterdam location in the Dutch projected Coordinate Reference System
.

\section{Results 1: Inspecting case study IFC models from practice}\label{sec:results1}

In the next subsections, the features of the BIM manually inspected are described as useful to check the considered regulations.

\subsection{Maximum height measurement}\label{sec:resgtmaxheight}

Maximum buildings height was measured on the delivered BIMs. 
The results, i.e.\ 103.47 m for \textit{Peak tower} and 100.46 m for the \textit{Terrace tower}, were consistent with the values reported by the municipality building permit documents, already issued following a traditional procedure.
Although such numbers are exceeding the prescribed limit of 100 m, they could be admitted anyway, since exemptions and derogation are foreseen in specific cases\footnote{Article 19.2 \url{https://www.ruimtelijkeplannen.nl/documents/NL.IMRO.0599.BP1054Waterstad-va01/r\_NL.IMRO.0599.BP1054Waterstad-va01.html\#\_19\_Algemeneafwijkingsregels} Accessed 4th November 2021}.
However, the non-compliance, as well as the possibility of obtaining a derogation, should be notified by the tool, and it could be implemented in future releases.

The condition to the exemption, in this case, consists in the fact that the part exceeding the 100 m limit is mainly occupied by installations.
By inspecting the BIM (considering the several models composing it all together) it is possible to verify this condition.
If the semantics of the elements, and/or of the related spaces, were attached consistently to the IFC model, the check could be completely automated.

\subsection{The subdivision of the BIMs in building storeys}\label{sec:storeysgroup}

According to the IFC schema, each \textit{IfcBuildingElement} is contained within an \textit{IfcBuildingStorey}, through the \textit{IfcRelContainedInSpatialStructure} property.
The entity \textit{IfcBuildingStorey} groups the elements belonging to the same floor, approximately from slab to slab.

First option to detect the building parts could be to assume that the blocks are composed by a consistent and continuous stuck of \textit{IfcBuildingStoreys}, seamless in facade.
However, this is not exactly what was found in the models, both because consecutive storeys can be slightly staggered, although belonging to the same building part, and because it is possible that variations in facade happen within the same storey, which would make the approach inaccurate.

\begin{figure}
	\centering
	\includegraphics[width=\linewidth]{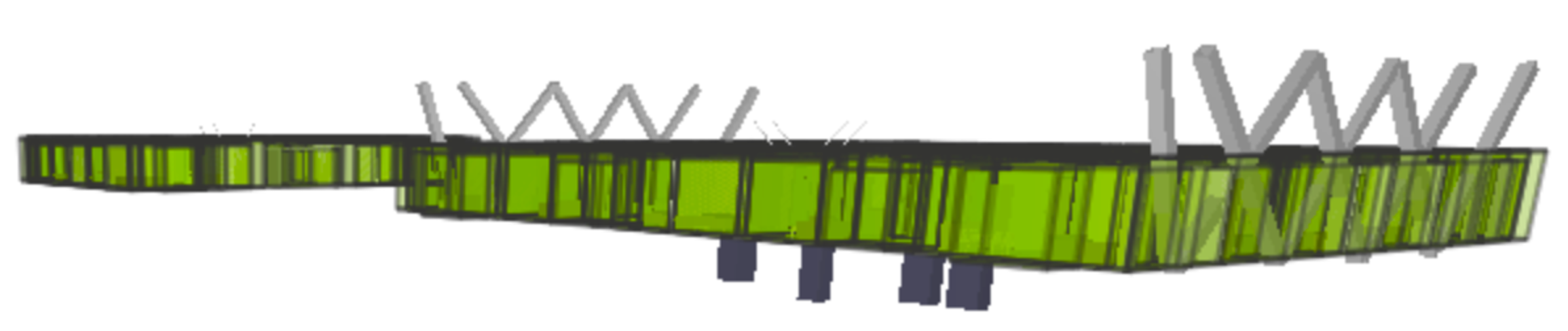}
	\caption{In \textit{Peak tower} BIM, In this picture of storey 13 it is possible to see how some elements necessarily extend for two floors, although not being a modelling error.}%
	\label{fig:stor2floors}
\end{figure}

For an initial simplified solution, the bounding box of each storey could be considered as reference.
But, again, the actual models present issues to implement such solution.
In particular, when checking the grouping consistency in the models, it is apparent how the assembling of entities in storeys often presents issues (as previously experienced also in other IFC models) \citep{noardo2021inspection} such as:

\begin{enumerate}
	\item One storey also includes elements belonging to different storeys;
	\item They contain solely a handful of isolated elements that should be included in another storey at similar elevation;
	\item The storey can contain elements that extend for two floors (e.g.\ \Reffig{fig:stor2floors}
	). The IFC standard suggests to break up walls at storey boundaries, but for other elements it is allowed that they span multiple levels (the \textit{IfcRelReferencedInSpatialStructure} is introduced to signify this relationship).
	It is not a modelling error, although the measurements coming from such grouping would not be accurate.
\end{enumerate}

To fix the cases 1 and 3, elements could be excluded from the grouping whether they do not fit within a buffer with respect to a reference height.
Reference could be, for example, the \textit{Elevation} attribute of the \textit{IfcBuildingStorey} or a statistical parameter calculated among the barycentres of the \textit{IfcBuildingElements} grouped in the storey.
For fixing the case n.2 (isolated elements grouped as one storey) other criteria could be used, such as a minimum threshold to the number of entities included in a valid storey.
Otherwise, comparisons between the bounding boxes of adjacent storeys could help in detecting such errors and exclude them.

An additional note is that the software exporting the IFC file usually considers all the elements at the same level as belonging to the same storey, therefore, the same floor of two different towers are grouped within the same \textit{IfcBuildingStorey}.

\subsection{Discontinuity detection criteria}

\begin{figure*}
	\includegraphics[width=0.65\linewidth]{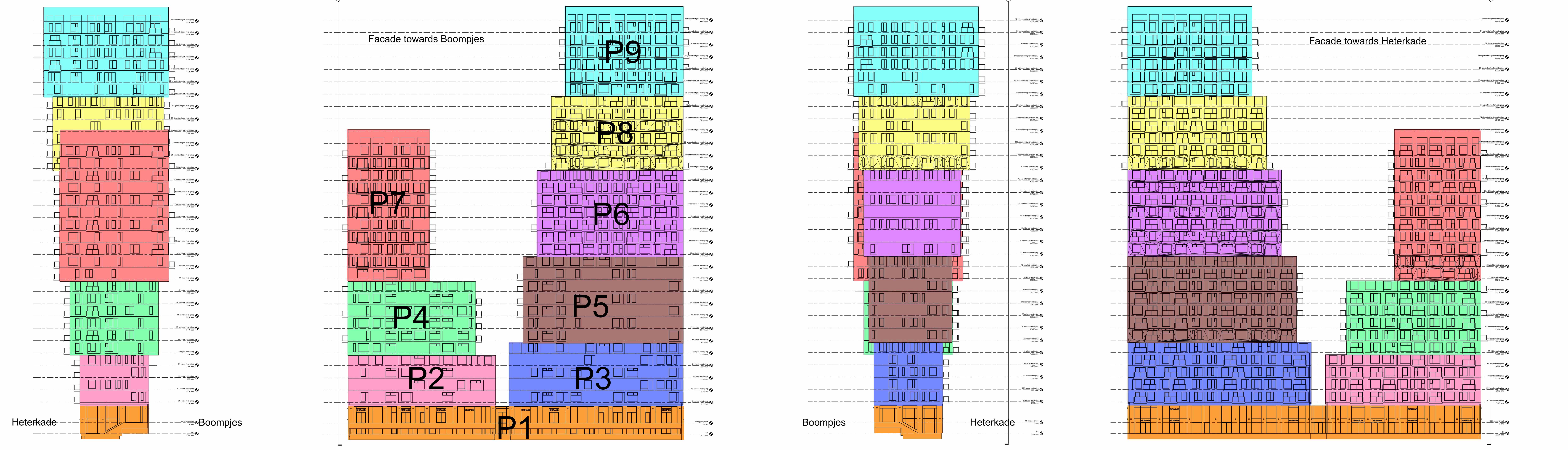}(a)
	\includegraphics[width=0.25\linewidth]{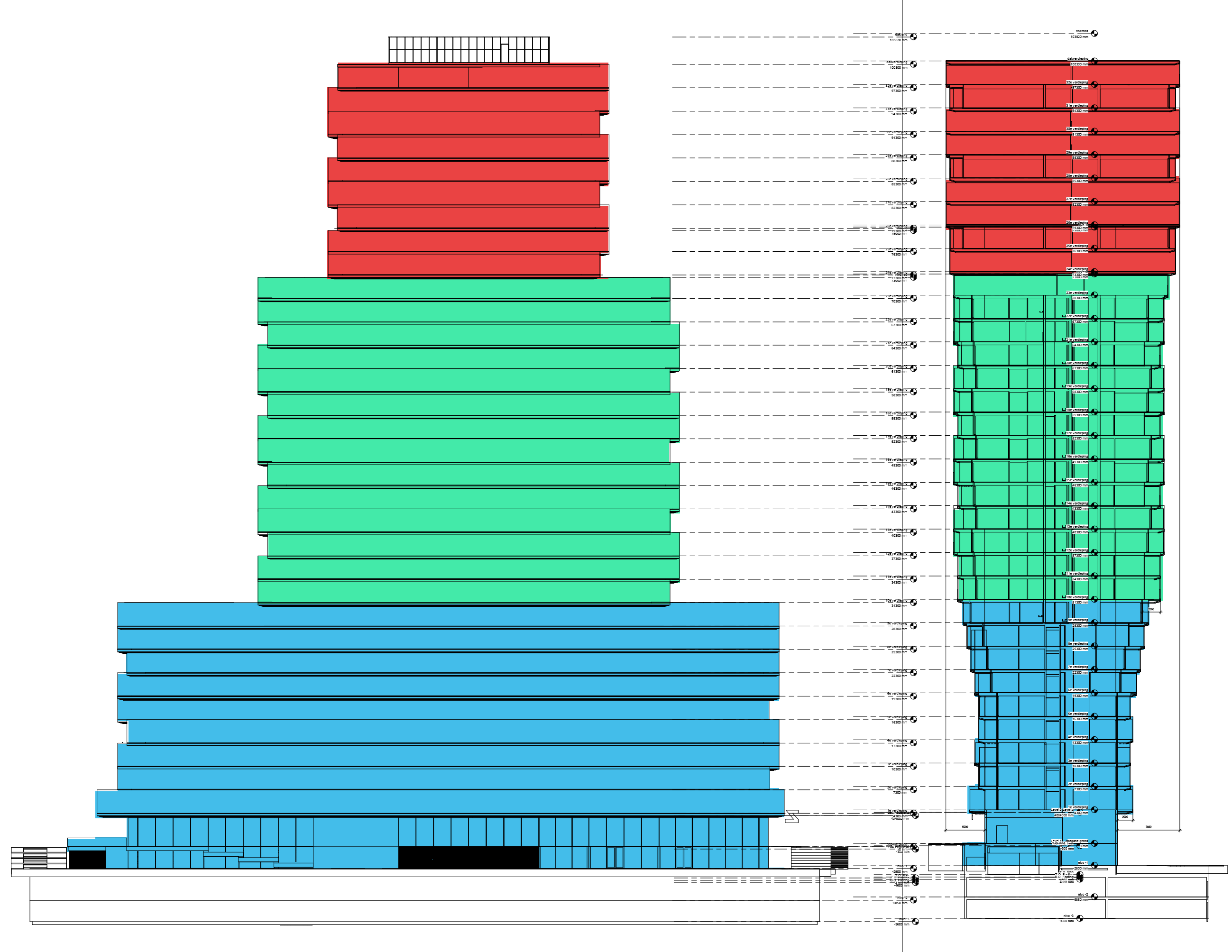}(b)
	\caption{Orthogonal projections of the facades of the Peak (a) and Terrace tower (b) BIMs segmented intuitively by human judgement.}%
	\label{fig:facadesseg}
\end{figure*}

The BIMs can be segmented according to the perception of vertical discontinuities by the human eye as in \Reffig{fig:facadesseg}, where it is possible to notice that the facades are articulated by the several elements, windows, and slight differences between the floors.
The dimension of protruding parts, which are not perceived as a discontinuity in the building shape by the human eye, were measured on the BIM, as well as the parts that do represent a discontinuity.
The result shows that a minimum change in extension of 5\% between consecutive floors determine the segmentation of the building in different parts, as it happens in the \textit{Peak Tower} as well.
This minimum change is supposed not to be distributed equally among the different sides of the building (e.g.\ 2.5\% per side), but to be the overall jump between two main vertical surfaces.
However, the generalized geometry of the building has to be taken into account: in the case of \textit{Terrace tower}, consecutive floors are staggered of a 5\% each other, but when considering larger series, they can be gathered in one block.
Therefore, accurate tools should automatically compare not only the coordinates of two consecutive storeys but they could group the floors based on their sequence and the variance of their extension with respect to the whole stack of storeys constituting the building.

The measurements of the heights of the segmented parts can identify which are part of the base and which ones belong to the top part.
Another criteria to detect the two parts of the \textit{Peak tower} could also be the separation of the part of building composed by only one contiguous part by the part split in the two towers.

\subsection{Measurement of the `base' height and the ceiling-slab-floor ensemble thickness in the facade}\label{sec:slabthickness}


To check the maximum height of the base (limit 17 m), the maximum height of the higher storey composing the base can be read, or the minimum height of the lower storey part of the top.
However, both could be inaccurate.
Alternatively, the height of the lower surface of protruding parts of the building, seen as a ceiling from outside the building can be inferred by considering the distance between this and the floor height as stored in the \textit{Elevation} attribute of the \textit{IfcBuildingStorey} \citep{20cibw78}.

\begin{figure}[H]
	\centering
	\includegraphics[width=0.6\linewidth]{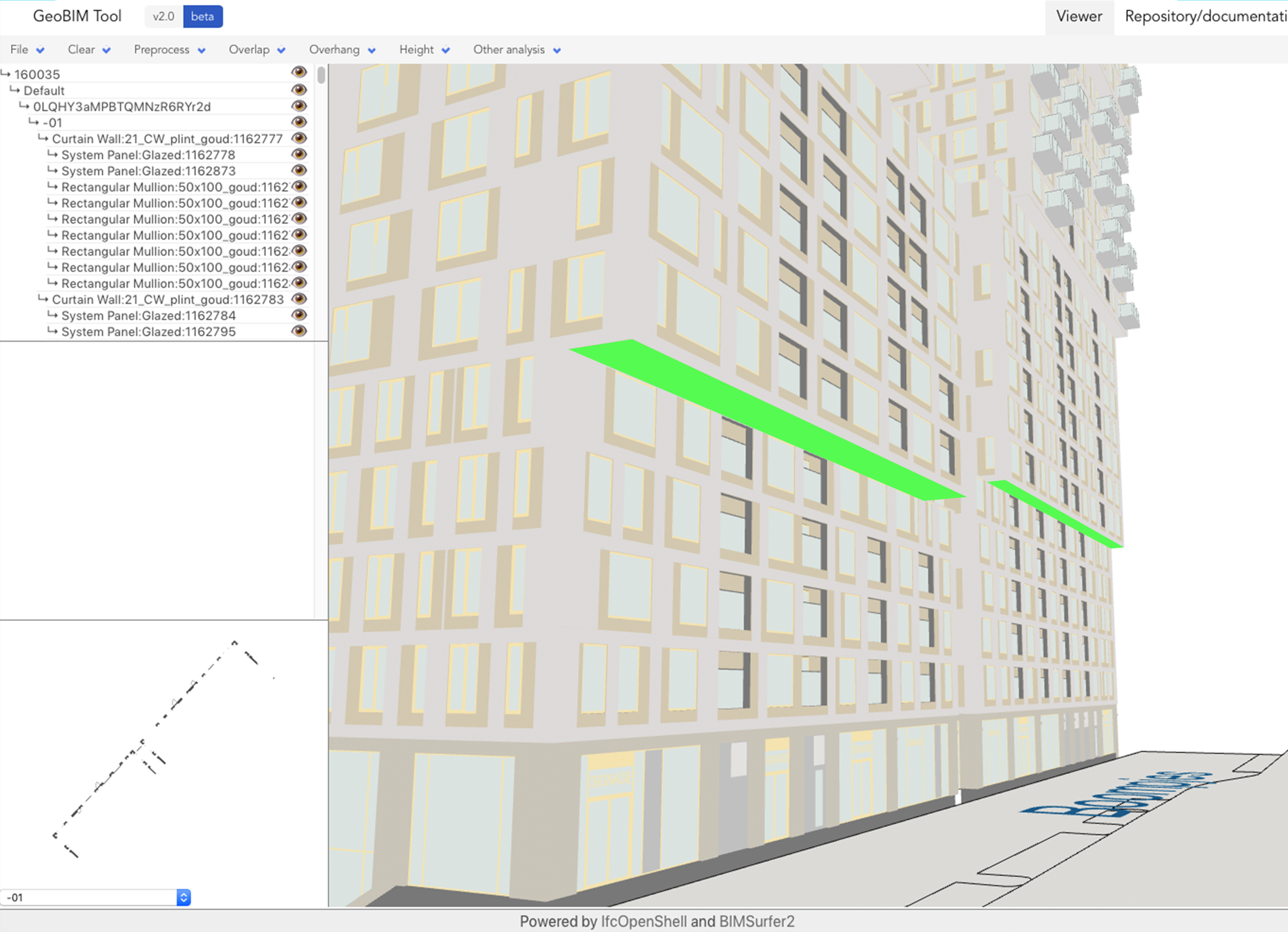}
	\caption{In \textit{Peak tower} BIM, in green, the ceiling to be measured to check the dimension regulation regarding the maximum height of the building `base'. The model is visualized in the web-based application developed as a follow-up of the initial investigation (\url{http://godzilla.bk.tudelft.nl/geobim-tool/viewer} Accessed 4th November 2021).}%
	\label{fig:protrudingceiling}
\end{figure}

However, the thickness of the ensemble ceiling-slab-floor, based on which such distance can be set, is not straightforwardly known.
For example, in the \textit{Peak tower} BIM, 26th floor, the elevation of the \textit{IfcSlab} is 81.61 m and of the lower surface protruding from the building (in green in \Reffig{fig:protrudingceiling}) is 81.02 m: the total thickness is therefore 0.59 m, while the slab is only 8 cm. 
Many design choices can influence such dimension (e.g.\ kind of construction and structural system, kind and amount of isolating materials, false ceilings, installation ducts or halls, specific facade systems etc.).

Therefore, it would be safer and easier to measure such height from the ground reference, rather than rely on assumptions.
However, a very rough approximation between 0.6 and 0.5 meters below the \textit{Elevation} attribute value of the lowest \textit{IfcBuildingStorey} being part of the `top' part of the building could provisionally work.

\subsection{Storing information in different IFC disciplinary models}\label{sec:distributedModel}

One BIM is usually composed by different disciplinary models (architectural, structural, installations, etc.), stored in different IFC files.
For consistency, the recommendation is to consider all such IFC models together for the calculations.
It is not possible to rely on only one of them since often the elements that a building design layman would consider as represented together are instead distributed in different models.
For example, structural walls could belong to the structural model, without the architectural model including them. 
Therefore this last one could not be used by itself to compute paths or spaces.
Moreover, structural (or installations) elements could extend until the facade.
Using only one of the models may imply excluding relevant information that is available in one of the other models.

\subsection{Storing information in the IFC models: semantics (in)accuracy}\label{sec:semantics}

The consistent labeling of objects and the filling of useful attributes is also critical.
The semantics in the models are compliant with the IFC standard.
However, semantics could be differently assigned, either by the user during the design or by the implementation or mapping tables between the IFC schema and the internal data model of the software.
Such inconsistencies could be due to an actual flaw, but they could also represent different interpretations of the IFC schema, in the cases where clear constraints or definitions are not present.

\begin{figure*}
	\centering
		\includegraphics[width=0.75\linewidth]{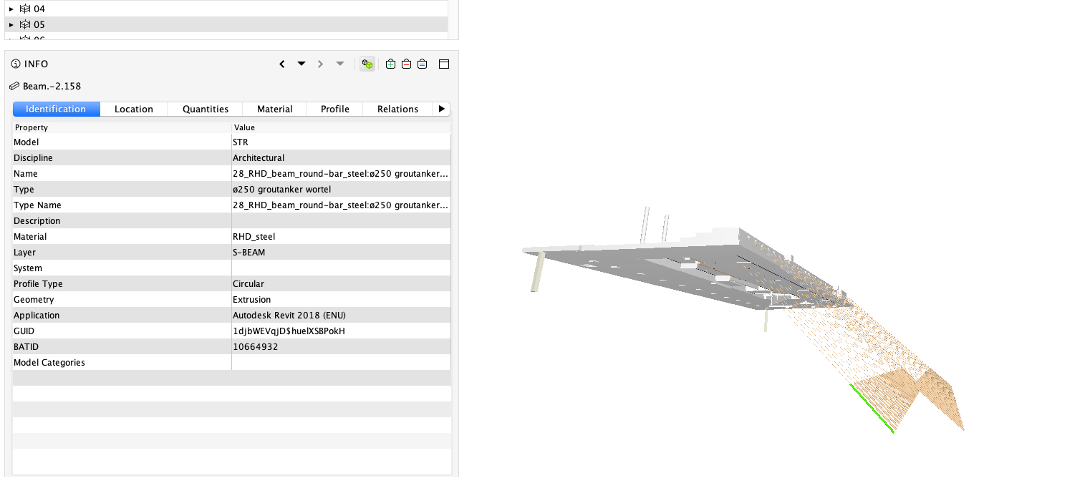}
	\caption{In \textit{Peak tower} BIM, an examples of semantics inaccuracies. The properties of the green object are listed in the tables on the left.}%
	\label{fig:terracewrongbeams}
\end{figure*}

According to the inspection of the two models in the case study, it is not wise to blindly trust the semantics as stored in the IFC model.
An example is the representation of balconies in the \textit{Peak tower} BIM\@: with the slabs being \textit{IfcWalls} (of course inconsistent representation) and an \textit{IfcRailing} in its correct place. 
In other cases, some structural elements are labeled as beams but probably have a different function, since their main extent is vertical.
This is one case where the plain translation of the label can be misleading (e.g.~in \Reffig{fig:terracewrongbeams}): a tie-rod can be called tie-'beam', but the structural function with respect to a normal beam is different and can possibly give issues when using the model for structural computations.

In addition, it is possible to notice that the elements to be counted in the building dimensions measurement could be various (including installations, chimneys, or \textit{IfcBuildingElementProxy}, the general IFC entity to represent whatever is not alternatively defined).
This fuzziness makes the elimination of any entity risky (with only possible exceptions of \textit{IfcOpenings} and \textit{IfcFurniture}).

At present, it is hard to rely on either the \textit{ifcType} assigned to objects or any further feature that could better define the role and function of each element such as the attributes \textit{'loadbearing'}, \textit{'material'}, \textit{'isExternal'} or the geometry itself.
In fact, tools could be programmed in order to use those elements to infer the correct IFC class of each object and fix the model, within an automatic or semi-automatic procedure still foreseeing some user interaction.
However, those features are irregularly used as well, and therefore an extensive work would be necessary to regulate the use of IFC, including attributes, explicit relationships and geometries, and, on the other hand, to develop a tool able to solve the remaining inaccuracies.
More advanced techniques, such as machine (supervised) learning could also be considered in future to tackle the same issue.
However, it is not possible at the moment, due to the lack of a sufficient sample of reference models to be used to train the algorithms as well as the lack of explicit agreements about the suitable use of the IFC data model.

\subsection{Intersecting elements in the BIM}\label{sec:overlappingelements}

\begin{figure}[H]
	\centering
	\includegraphics[width=0.7\linewidth]{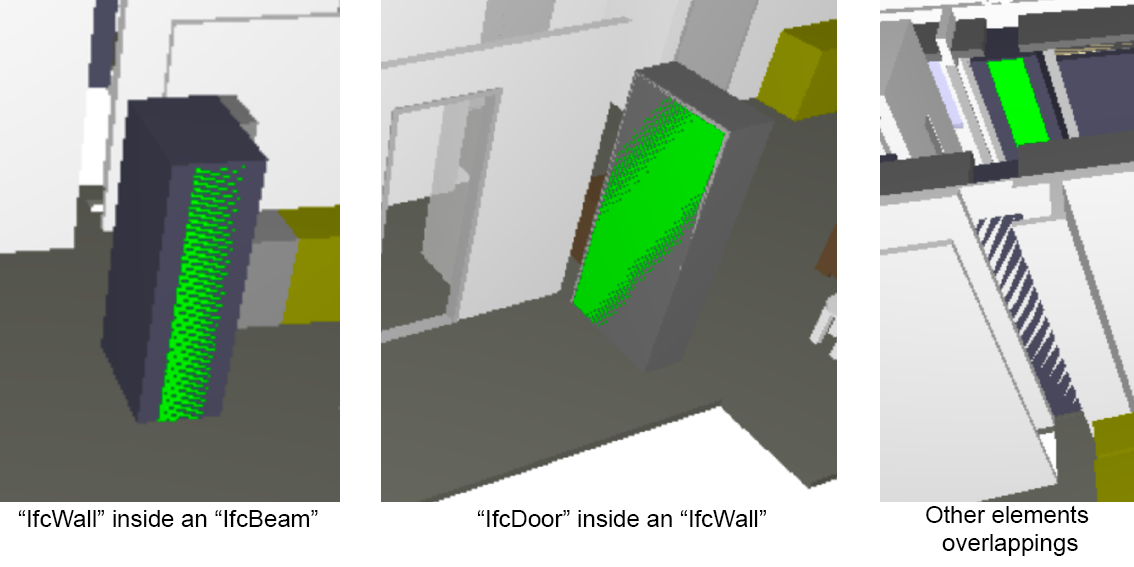}(a)
	\includegraphics[width=0.25\linewidth]{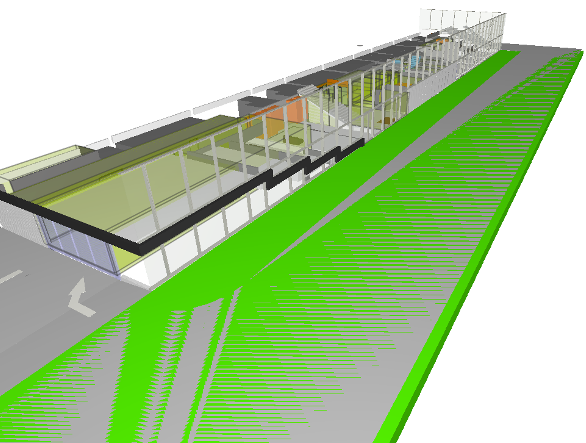}(b)
	\includegraphics[width=0.25\linewidth]{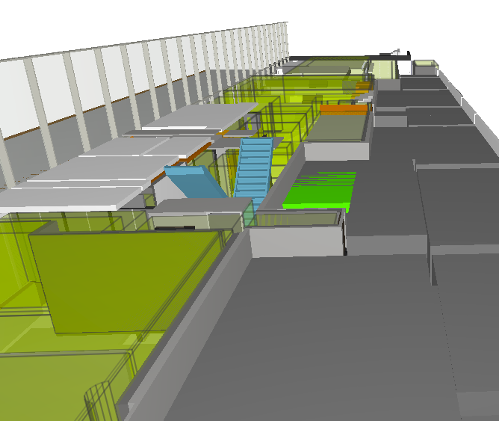}(c)
	\caption{(a) In \textit{Peak tower} BIM, some examples of elements overlap. (b,c) In \textit{Terrace tower} BIM, some examples of elements overlaps: in (b) two slabs, in (c)  many elements, modeling the bikes parking spaces and represented by means of \textit{IfcBuildingElementProxy}. The bright green is the selected element, that is represented as superimposed to another one, so that two faces corresponds (the stripes and blurry colors are shown in that case)}%
	\label{fig:TerraceElementsOverlap}
\end{figure}

A further issue is about elements that intersect each other (walls and beams, walls and slabs, walls and doors, as a few examples) (e.g.\ 
\Reffig{fig:TerraceElementsOverlap}). 
This can be due not only to inaccuracy in the modeling phase, but also to the implementation of export algorithms in software.
The consequence is that one cannot always rely on the geometry present in the IFC model, besides limiting possible automatic processing.

\subsection{Use and modeling of Spaces and \textit{IfcSpaces} in the BIM}\label{sec:spaces}

Both the \textit{Peak} and the \textit{Terrace tower} models present an interesting modeling of \textit{IfcSpaces}, potentially helpful both for the extraction of the building envelope geometry and for checking the parking regulation.
A redundant but comprehensive part-of hierarchical modeling of spaces is provided: 

\begin{description}
 \item[Space 1]\label{sp1} One space per floor includes the whole storey area, excluding only the boundary slabs  (\Reffig{fig:spaceStorey}(a));

\item[Space 2]\label{sp2} includes the same than \textit{Space1}, but excludes the vertical connection spaces and walls bordering with the exterior of the building (\Reffig{fig:spaceStorey}(b));

\begin{figure}[H]
	\centering
	\includegraphics[width=0.43\linewidth]{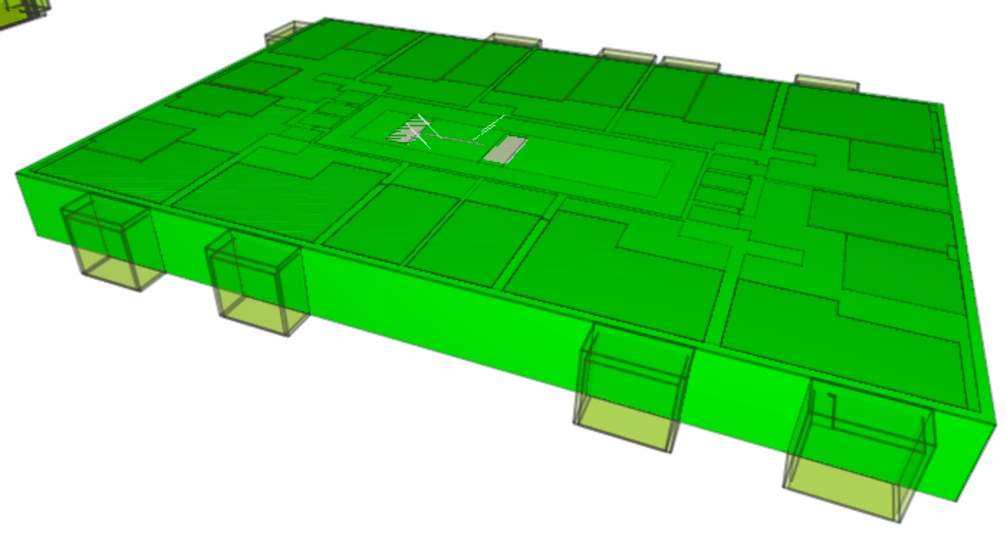}(a)
	\includegraphics[width=0.43\linewidth]{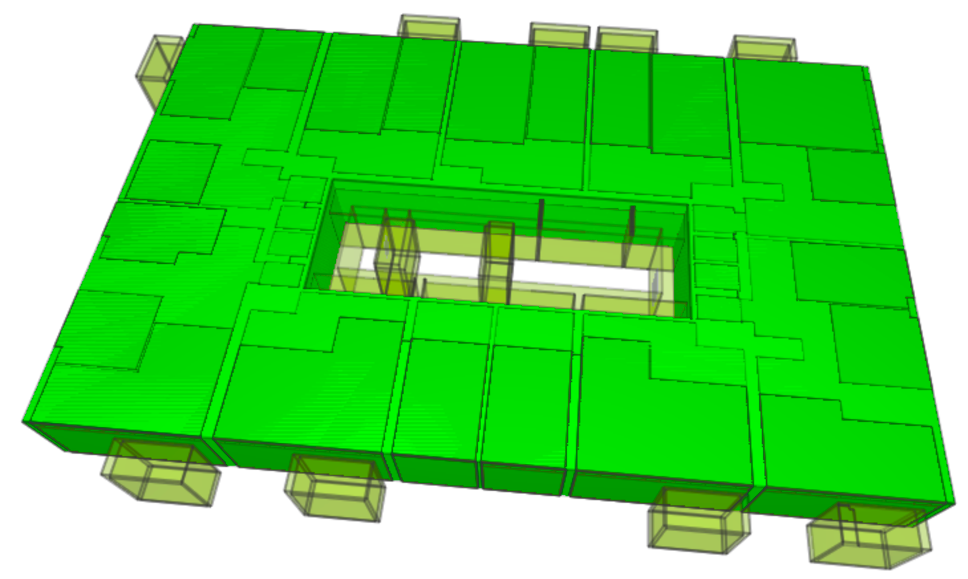}(b)
	\caption{Example of floor spaces in \textit{Peak tower} BIM\@. In (a)(Space1) it is a box including the whole storey, gross space inside the exterior walls and slabs. Attribute \textit{Type} `BVO' (\ie\ gross floor area). In (b) (Space2) it is a box including the whole storey, except for the vertical connection spaces and installations at the center of the building and the external walls.}%
	\label{fig:spaceStorey}
\end{figure}

\item[Space 3]\label{sp3} Include single apartments (e.g.\ \Reffig{fig:spaceApart});

\begin{figure}[H]
	\centering
	\includegraphics[width=0.43\linewidth]{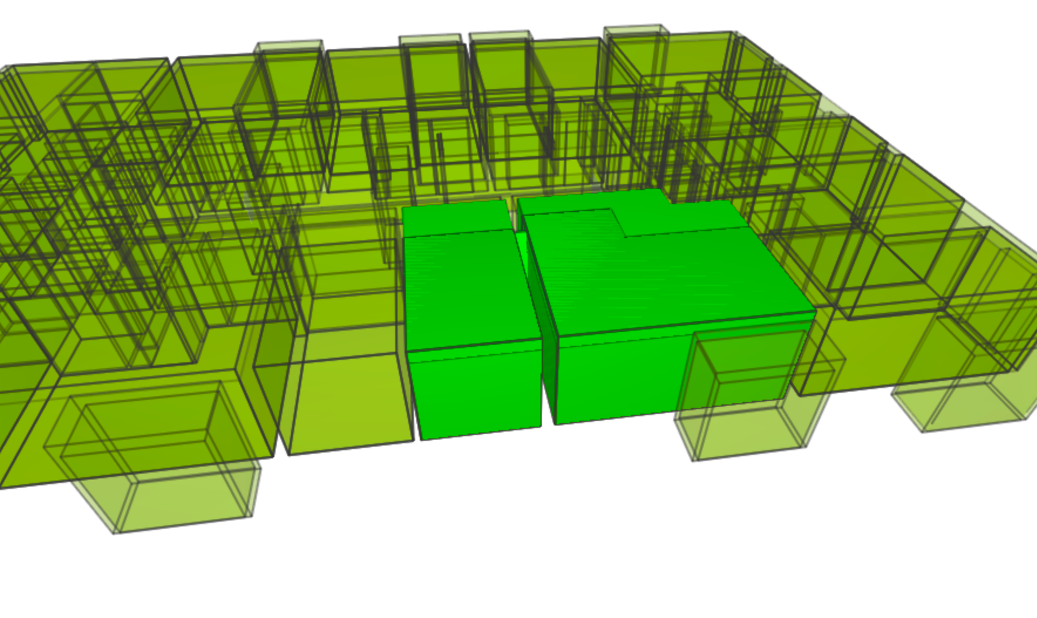}(a)
	\includegraphics[width=0.43\linewidth]{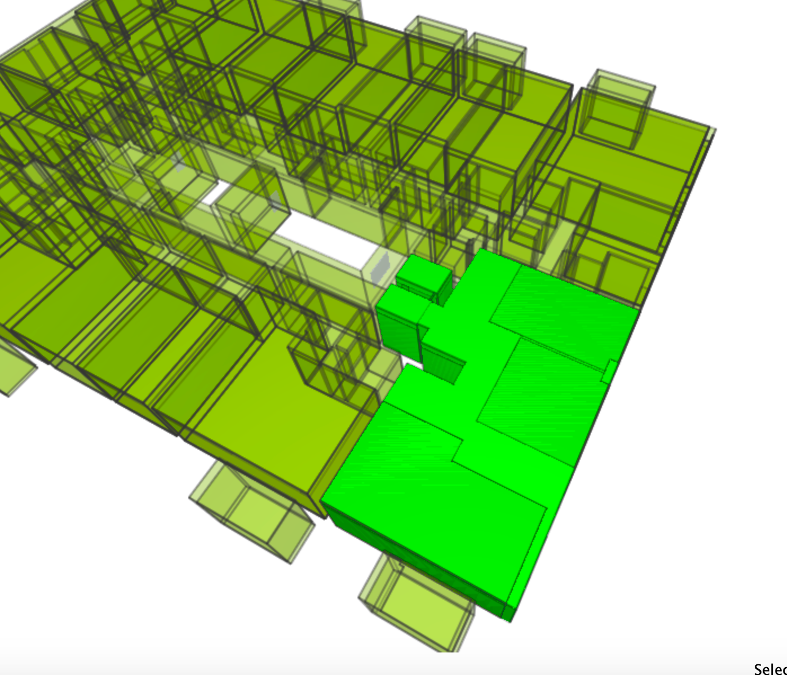}(b)
	\caption{Example of apartment-level spaces\@. (a) \textit{IfcSpace} with attribute \textit{Type} having value `B50 – 70m2' (area = 65.57 m$^2$), representing only the interior of rooms (attribute \textit{interior} `True'). (b) \textit{IfcSpace} with attribute \textit{Type} having value `E70 - 140m2' (area=82.68m$^2$).}
	\label{fig:spaceApart}
\end{figure}

\item[Space 4]\label{sp4} Includes the single rooms and balconies. 


\end{description}

Looking at the spaces in detail, further unclear issues can be noticed.
For example in the modeling of apartments (\Reffig{fig:spaceApart}), some of the internal walls are included and some others are not.
Usually they do not include the thickest walls and the facade.
Maybe this is due to the generation of spaces starting from one of the models (e.g.\ the structural one, or the architectural one), which make the modeling software refer to the information present there.
Moreover, the same spaces (especially Space 4) are represented with many overlapping volumes, all starting from the slab, sometimes with different heights (\eg\ until the top of the ceiling or at the walls height, etc.) but sometimes just repeated, which is unexpected, and would deserve to be investigated further with the help of designers.
In other cases, \textit{IfcSpaces} with the same label (stored as name and/or as type) can represent different spaces. 
On the contrary, redundant \textit{IfcSpaces} representing the same volume sometimes are labeled differently, but without explicit rules.
Finally, there are different spaces intersecting, but it is not possible to understand the different meanings of them and consequently if and why are they allowed to overlap.



Such a representation of \textit{IfcSpaces} could be indispensable to more easily process the BIM automatically, although such a modeling is not a standard prescription nor official best practice (IFC foresee the representation of apartments as described in Section~\ref{sec:apart}).
In the case study models, either the attribute \textit{Name} or \textit{type} usually store the main semantics and codes.\footnote{
In the \textit{Peak tower} BIM IFC models, the spaces describing the apartments report values like ``Type B 50--70 m2'' and similar 
under the `\textit{Name}' attribute, which cluster the apartments based on their floor area.
This was helpful to detect the \textit{IfcSpaces} describing a whole apartment to calculate the reference areas for parking places calculation.
In the \textit{Terrace tower} BIM, they are explicitly labeled with the \textit{Name} and \textit{Type} ``Appartment''.}
However, there is no clear documentation helping in their interpretation, nor a standard classification is followed.

\begin{figure}
	\centering
	\includegraphics[width=0.43\linewidth]{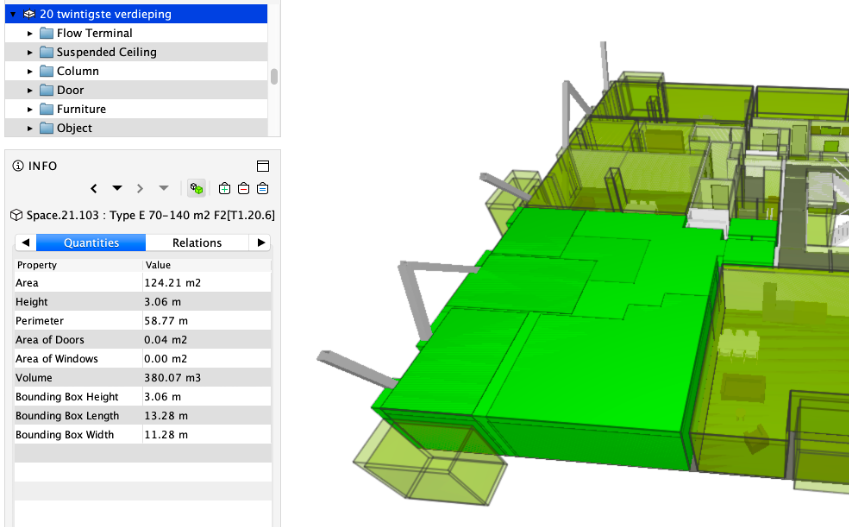}(a)
	\includegraphics[width=0.43\linewidth]{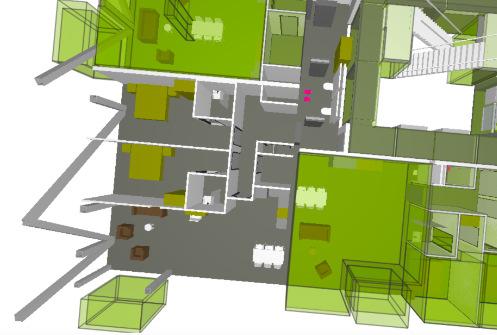}(b)
	\includegraphics[width=0.43\linewidth]{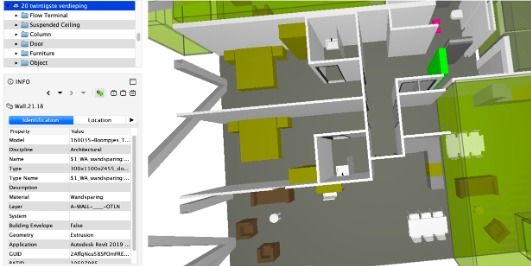}(c)
	\caption{Example of one apartment represented in the \textit{Peak tower} BIM\@: (a) The \textit{IfcSpace} enclosing it\@; (b) How it appears in the architectural model, it is possible to notice the absence of (structural) walls\@; (c) The bright green element is an \textit{IfcWall}, but it should be a door.}%
	\label{fig:spaces}
\end{figure}

Alternatively, apartments areas could be extracted starting from the complete model of interiors, according to the paths allowed by the building elements dispositions.
However, challenges would be involved (\Reffig{fig:spaces}):

\begin{description}
	\item[Issue 1] Parts missing (e.g.\ structural elements not in the architectural model);
	\item[Issue 2] Semantics wrongly defined (e.g.\ walls instead of doors prevent a possible algorithm to understand path/non path).
\end{description}

\subsection{Representation of parking places}\label{sec:inspectionParkingplaces}

\begin{figure*}
	\centering
	\includegraphics[width=0.4\linewidth]{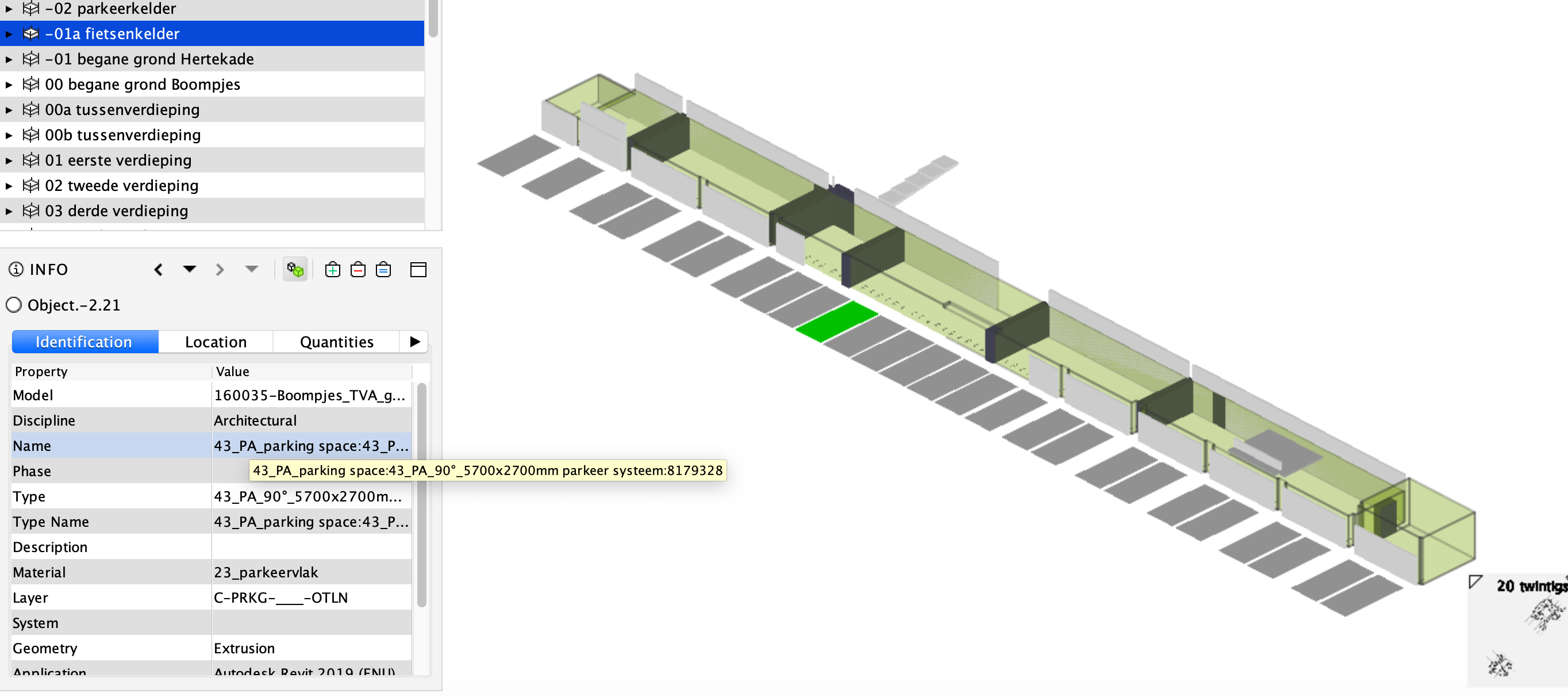}(a)
	\includegraphics[width=0.4\linewidth]{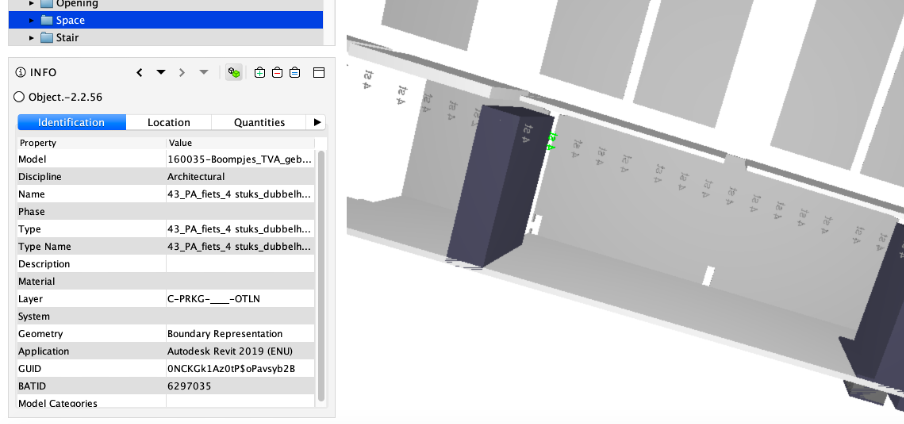}(b)
	\caption{Parking places for cars (a) and for bikes (b) in the \textit{Peak Tower} BIM.}%
	\label{fig:carparkingsBoomp}
\end{figure*}

In the \textit{Peak tower} BIM, the parking places are represented by means of \textit{IfcBuildingElementProxy}, with \textit{Pset\_ProductRequirements} $\rightarrow$ Category= Parking (\Reffig{fig:carparkingsBoomp}).
It is therefore possible to count them: they are 57, distributed on the two underground floors.

The bike parking places are represented as \textit{IfcSpace} with Name ``\textit{fietsenstalling}''.
In addition, text is added to the model, reporting the number of bike places hosted in an area, but it is not stored as any attribute to be automatically processed, besides being a very undefined representation even for a human check (and the designer should be trusted, since the area is not drawn specifically).

In the \textit{Terrace tower} architectural IFC model, one space includes the whole storey, labeled as ``\textit{parkeren}'' and another overlapping one labeled as ``\textit{stallingruimte (garage)}'', inclusive of both the area of parking places and the path for cars movement.\footnote{For more extensive description and images see \url{https://3d.bk.tudelft.nl/projects/rotterdamgeobim_bp/Rotterdam2DBP.pdf}}%
However, neither the slab nor walls dividing the space are represented, as probably included in the structural model.
In addition, many boxes, (as \textit{IfcBuildingElementProxy}) intended to represent parking spaces for bikes are overlapping, and therefore not reliable.
Thus, it is not possible to proceed with the counting of parking places and check the regulation neither automatically nor manually by starting from the information stored in the IFC models.

\section{Results 2: the implemented checking tool demonstrator}\label{sec:restool}

The implemented tool was developed to extract as much information as possible from the BIM overall geometries, without relying on the IFC geometries of the single elements and only in a few cases on the stored semantics.
The open source code and executable can be downloaded at \url{https://github.com/tudelft3d/GEOBIM\_Tool}.
Once downloaded, the\ .exe file can be launched directly, stored in the folder where the references files and folders are hosted (the\ .py files, the\ .yaml files specifying the parameters to use, and the folders where the results will be stored).
An interface will allow the user to open, load and visualize the IFC file and use the tools.
Some documentation, including a video tutorial, can also be found in the GitHub repository.

\subsection{Maximum height calculation by the implemented tool}\label{sec:maxh}

The tool measures the maximum height as the distance between the top of the BIM (the highest $z$ value of the uppermost IFC element) and the bottom of the ground floor, intended as the value of the \textit{Elevation} attribute of the \textit{IfcBuildingStorey} representing the ground floor.

In order to consider the possible differences between aspect models, it is recommended to run the same tool on all the models and consider the highest value, since the starting point (\textit{IfcBuildingStorey.Elevation} of the ground floor) should be consistent through all the models.

One more check should be necessary, considering the road elevation adjacent to such entrance, instead of the entrance itself.
This will be something to be addressed within future work, when working with actual Geo and BIM integration.

\subsection{Storey overlap calculation algorithm}\label{sec:algovcalc}

This algorithm extracts the storeys profiles and calculates their reciprocal overlapping.
From the obtained result, the groups of storeys composing the base and the top part are identified and their overlap computed.

\begin{figure}[H]
	\centering
	\includegraphics[width=0.8\linewidth]{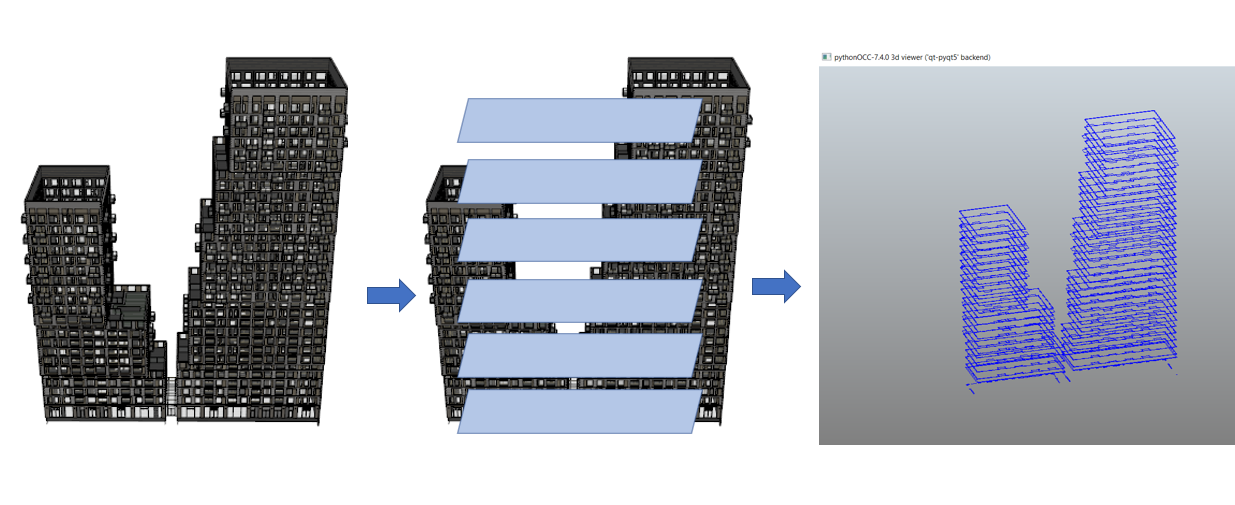}
	\caption{Section cut of each building storey with a horizontal plane.}
	\label{fig:sectioncut}
\end{figure}

A reference storey, such as the ground floor, must be selected, to be reference for the comparison to the other floors.
Second, each building storey is cut by a horizontal plane whose height is set starting from the \textit{IfcBuildingStorey.Elevation}\footnote{\url{https://standards.buildingsmart.org/IFC/RELEASE/IFC4\_1/FINAL/HTML/schema/ifcproductextension/lexical/ifcbuildingstorey.htm}} 
(Figure~\ref{fig:sectioncut}).

\begin{figure}[H]
	\centering
	\includegraphics[width=0.8\linewidth]{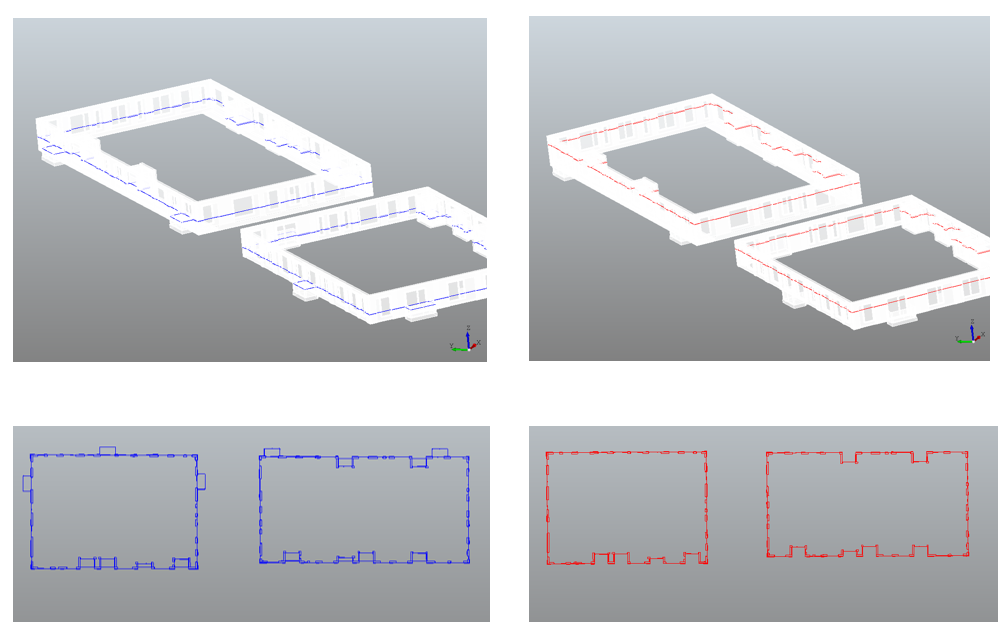}
	\caption{Result of the extraction of the storeys profile, using two different heights as input parameters (blue: balcony included, red: balcony excluded)}%
	\label{fig:balconycut}
\end{figure}

The cutting height has impact on the inclusion of some facade elements of each floor, such as balconies (\Reffig{fig:balconycut}). 
Users can define the cutting height with respect to \textit{IfcBuildingStorey.Elevation} based on their own requirements.

\begin{figure*}
	\centering
	\includegraphics[width=1\linewidth]{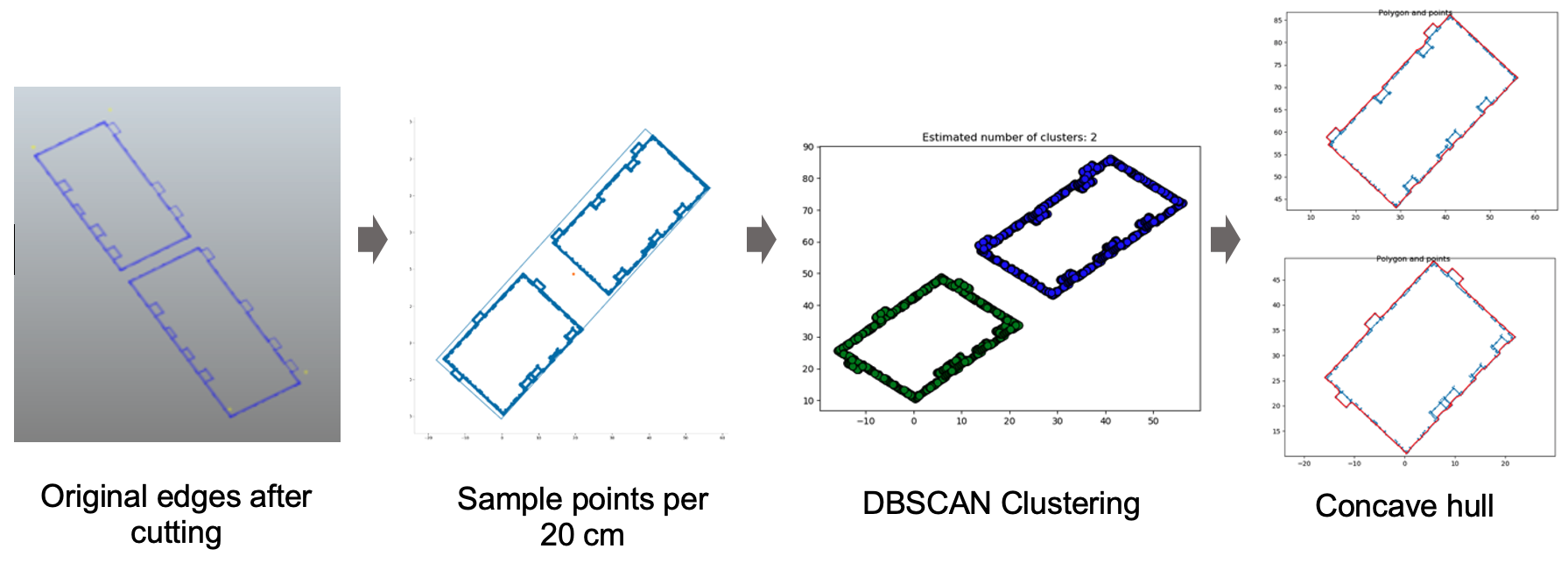}
	\caption{Overview of the automatic polygon detection from the cutting result}%
	\label{fig:autopoly}
\end{figure*}

The intersection with the building element geometries of each storey is stored as 2D polygons, from which the footprints are reconstructed.
First step is to sample points every 20 cm in each edge of the polygon.
Second, clustering those points by using the Density-based spatial clustering of applications with noise (DBSCAN) algorithm~\citep{Martin1996}.
Finally, the concave hull of each point cluster is generated through the concave hull algorithm based on the k-nearest neighbors by \citet{moreira2007} (\Reffig{fig:autopoly}): 
A starting point is selected, such as the point with the minimum Y value, considered as the `current point'.
The $k$-nearest neighbors of the current point are collected.
The angles between the current point and neighbors are calculated. The one with the largest right-hand turn angle measured from the horizontal line is included in the concave hull and becomes the new current point.
The process is repeated until the start point is selected again.

At this point, the overlap percentage between the polygon representing the footprints of each floor and the polygon of the footprint of the reference storey is calculated by using polygon intersection.
Table~\ref{table_overlap} gives an example of the results of the overlap algorithm run on the \textit{Peak tower} BIM\@, with the ground floor (manually) chosen 
as the reference storey.

\begin{table}
	\centering
	\caption{Overlap percentage of \textit{Peak tower} BIM, base: ground floor}\label{table_overlap}
	\begin{tabular}{|l|l|l|l|}
		\hline
		Floor name   & \begin{tabular}[c]{@{}l@{}}Overlap (\%)\end{tabular} & Floor name & \begin{tabular}[c]{@{}l@{}}Overlap (\%)\end{tabular} \\ \hline
		ground & 100                                                              & 17th       & 67.9                                                             \\ \hline
		1st    & 94.1                                                             & 18th       & 67.9                                                             \\ \hline
		2nd    & 93.8                                                             & 19th       & 67.9                                                             \\ \hline
		3rd    & 95.3                                                             & 20th       & 64.0                                                             \\ \hline
		4th          & 93.9                                                             & 21st       & 63.6                                                             \\ \hline
		5th          & 88.9                                                             & 22nd       & 63.7                                                             \\ \hline
		6th          & 85.9                                                             & 23rd       & 39.2                                                             \\ \hline
		7th          & 85.9                                                             & 24th       & 39.2                                                             \\ \hline
		8th          & 86.6                                                             & 25th       & 39.3                                                             \\ \hline
		9th          & 86.2                                                             & 26th       & 35.0                                                             \\ \hline
		10th         & 85.9                                                             & 27th       & 35.0                                                             \\ \hline
		11th         & 72.3                                                             & 28th       & 35.1                                                             \\ \hline
		12th         & 72.0                                                             & 29th       & 35.0                                                             \\ \hline
		13th         & 67.9                                                             & 30th       & 35.2                                                             \\ \hline
		14th         & 67.9                                                             & 31st       & 35.0                                                             \\ \hline
		15th         & 67.9                                                             & 32nd       & 35.0                                                             \\ \hline
		16th         & 67.9                                                             &            &                                                                  \\ \hline
	\end{tabular}
\end{table}

By setting a threshold in the percentage difference, it is possible to detect, either manually or by implementing an algorithm, at which floors the building can be cut in different parts.

If the base appears to include more than just the ground floor, the computation can be run again, by setting the reference storey consequently.
At that point, any overlapping percentage larger than 50\% related to the storeys on the top of the highest storey still part of the base would detect a potential non-compliance to the regulation.

\subsection{Overhang distance checking algorithm}\label{sec:algovdist}

\begin{figure}
	\centering
	\includegraphics[width=\linewidth]{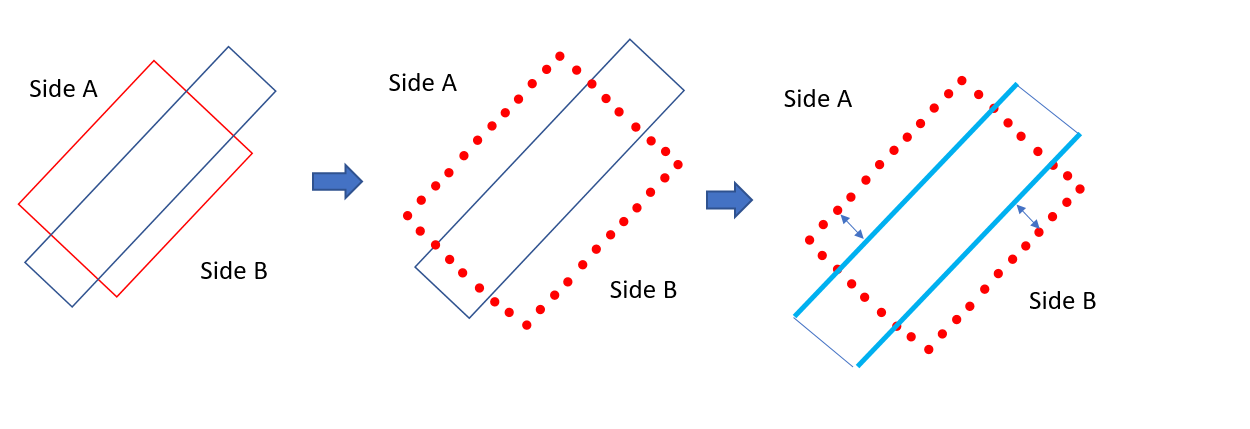}
	\caption{Overview of overhang distance calculation: blue polygon is the base floor footprint and the red one represents target floor; in the middle, red points are the vertex extracted from the geometries of the target floor \textit{IfcObjects}; on the right, the distances between each red vertex and the highlighted lines are calculated.}%
	\label{fig:overhang}
\end{figure}

The overhang distance calculation algorithm is implemented by using \textit{IfcOpenShell}~\citep{Thomas2020} and the pythonocc library~\citep{pythonocc2020}.
It consists of four steps (\Reffig{fig:overhang}):
selection of the base;
extraction of the vertex from all the \textit{IfcObjects} in the target floors using pythonocc; 
manual selection of the two highlighted lines of the base box, as the overhang calculation origins towards the two streets; calculation of the distance between each vertex and the selected lines.
Then maximum values are the overhang distances towards the two direction (\Reftab{tab:overhang}).

The selection of the facade lines in the two directions was manual.
Future developments should take advantage of the integration with geoinformation to compute such directions automatically.

\begin{table}
	\caption{Overhang result of the \textit{Peak tower} BIM\@. The balconies are included, but they can be excluded by changing parameters.}%
	\label{tab:overhang}
	\centering
	\begin{tabular}{|l|l|l|}
		\hline
		Floor name & \begin{tabular}[c]{@{}l@{}}Overhang distance \\ North, Hertekade\end{tabular} & \begin{tabular}[c]{@{}l@{}}Overhang distance \\ South, Boompjes\end{tabular} \\ \hline
		27th floor & 10.5 meter                                                                         &                                                                                   \\ \hline
		12th floor &                                                                                    & 6.4 meter                                                                         \\ \hline
	\end{tabular}
\end{table}

\subsection{BIM to GIS}\label{sec:togis}

\begin{figure*}
	\centering
	\includegraphics[width=1\linewidth]{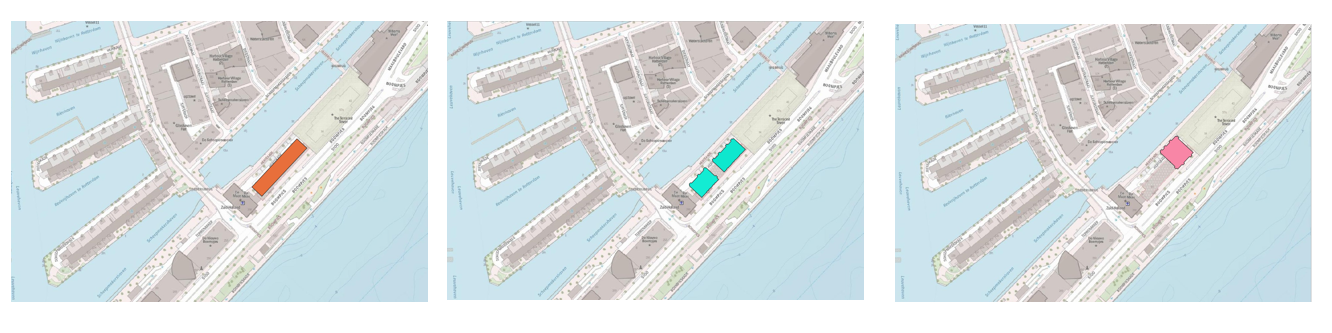}
	\caption{WKT geometry representation in QGIS of ground floor (in orange), 7th floor (in green) and 29th floor (in pink).}%
	\label{fig:wkt}
\end{figure*}

Although a complex and complete integration of BIM with geoinformation was not performed, the geometry delivered within the BIM is extracted and generalized properly by the algorithm described in Section~\ref{sec:algovdist} 
to be reused in the update of 2D maps.
In fact, either the georeferenced footprint of the building or the maximum outline of it, depending on the needs, can be saved as a Well-known text\footnote{\url{https://en.wikipedia.org/wiki/Well-known_text_representation_of_geometry}} files and easily added to the city map (Figure~\ref{fig:wkt}).

This allows a fast update of 2D maps whenever the building permit is issued.
In addition, the use of the footprint together with the heights 
would allow modeling LoD1 representations by means of quite simple processing.

\section{Results 3: Guidelines}\label{sec:resguidelines}

According to the lessons learned from the investigation of the data (Section~\ref{sec:results1}) 
and the implementation of the demonstrator (Section~\ref{sec:restool})
, we propose here some guidelines.

In order to support the integration with the geoinformation for automating the checks, some information (accurately georeferenced in the national projected CRS) should be preliminary provided to designers:

\begin{enumerate}
	\item Zoning parcels, stored with explicit attributes and unique identifiers related to the regulation text. For example, the regulation should be written as" ``in parcel with code\ldots a rule is valid'', instead of ``in Boompjes 50-59'' as it is written now, in the dimension regulation that we considered.
	\item City elements useful to assess the building in its context, such as terrain model, streets (useful in this case to detect the direction ``towards'' them, on which the regulation depends) etc. Each element should have unique consistent identifiers, better if written as a meaningful code for human interpretation, besides being a random string, to which the regulations could refer.
	\item Accurately measured network of reference points with 3D coordinates (with approximated reciprocal distance of 50 m), to be used as control points in the georeferencing (and as check points to assess its quality in the most difficult cases).
\end{enumerate}

Guidelines to planners are:  leave as little ambiguity as possible in the regulation text; quantify the involved parameters as much as possible; and make reference to other information by means of formal identifiers (\eg\ the code of the parcel or the streets).
Those are premises to any further work about systematic formalization of regulations.

In order to produce suitable IFC models for automatic processing, some guidelines to BIM modelers are listed in \Reftab{tab:glbim}.
More specific instructions and constraints should be developed in collaboration with the modeling practice field and evolve in clear and formal Information Delivery Manuals (IDM) to explain the detailed information required in the models.

Guidelines and tools to help controlling the IFC file are also essential for storing the georeferencing properly \citep{uggla2018geographic,jaud2020georeferencing,noardo2020tools}, without which it is impossible to proceed to the integration with the city context.

\begin{table*}
	\centering
	\begin{tabular}{|m{0.5cm}|m{8cm}|m{6cm}|}
		\hline
		\textbf{N.} & \textbf{Guideline} & \textbf{Aim}       \\ \hline
		1 & Group the objects external to the building in the \textit{IfcSite}, separating them clearly from the elements which are part of the building (in \textit{IfcBuilding}).
		&  Avoid disturbance in the computation of the geometry of the building such as exterior envelope or bounding boxes.
		\\ \hline
		2 & Ensure the correct and consistent grouping of \textit{IfcBuildingStoreys} avoiding and fixing misassignments (\ie\ inaccuracies pointed out in Section~\ref{sec:storeysgroup}).
		& Avoid disturbance in the computation of envelope and bounding box of the building, as well as the measurements of dimensions such as maximum and minimum heights.      \\ \hline
		3 & Use correct semantics and avoid using \textit{IfcBuildingElementProxy}.
		& Support any automatic processing.    \\ \hline
		4  &  Model the elements properly, in order to avoid intersections.
		& Support any automatic processing. \\ \hline
		5 & Georeference the IFC model preferably with minimum \textit{LoGeoRef 30} (see \citet{clemen2019level}).
		& Support integration with geoinformation.  \\ \hline
		6  & Model \textit{IfcSpaces} consistently and with explicit criteria
		& Help conversions and procedures involving the measurement of volumes and surfaces.     \\ \hline
		7 & Model building units or zones as prescribed by the standard 
		or by alternative methods, such as by \textit{IfcSpaces} or \textit{IfcZones}\footnote{New in IFC version 4.}
		, on the condition that they are explicit. Functions at the suitable level of hierarchy should be assigned consistently, preferably according to known classifications.
		& Support the detection and interpretation of spaces and related calculations. \\ \hline
		8  & Model the car and bike parking places explicitly (geometry and attributes) according to the prescription by buildingSMART of using \textit{IfcSpaces} in association to specific properties\footnote{\eg\ see \url{https://standards.buildingsmart.org/IFC/RELEASE/IFC2x3/FINAL/HTML/ifcproductextension/lexical/ifcspace.htm} with \textit{Pset\_SpaceCommon}: common property set for all types of spaces; \textit{Pset\_SpaceParking}: specific property set for only those spaces that are used to define parking spaces by \textit{ObjectType} = `Parking'; \textit{Pset\_SpaceParkingAisle}: specific property set for only those spaces that are used to define parking aisle by \textit{ObjectType} = `ParkingAisle'}.
Their effectiveness could be possibly discussed with architects and developers of the tools supposed to process the resulting BIM\@.
		& Support automation of parking checks.  \\ \hline
		9 & Register all the IFC models composing a BIM in a same (at least local) system (quite often it is current practice already). & Support automatically processing all the models consistently.\\ \hline
	\end{tabular}
\caption{Initial guidelines to produce suitable IFC files}%
\label{tab:glbim}
\end{table*}

\subsection{The representation of hierarchical building composition units according to IFC}\label{sec:apart}
 
\begin{figure}[H]
	\centering
	\includegraphics[width=0.7\linewidth]{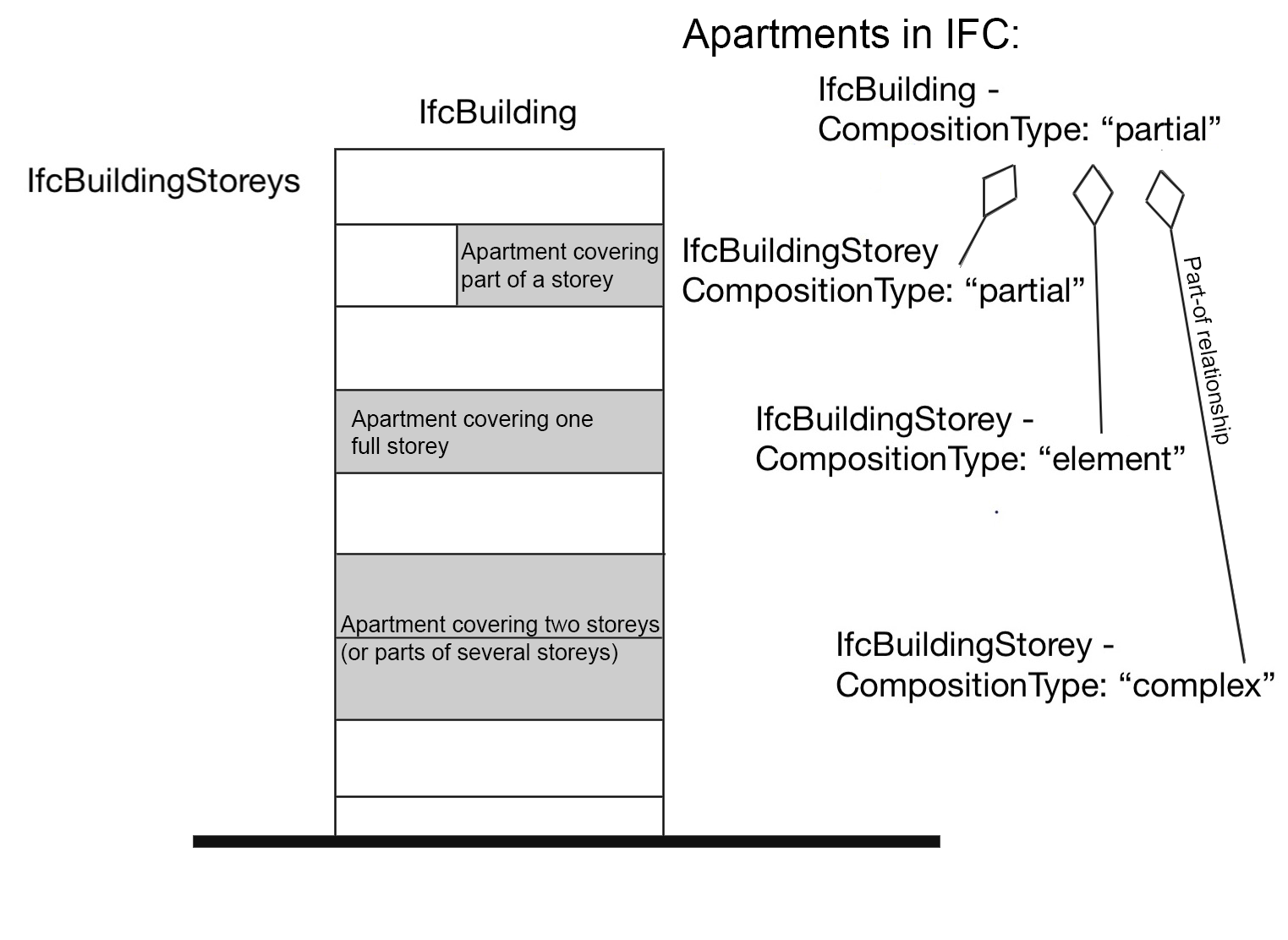}
	\caption{Schema of possible representation of apartments according to the IFC schema.}%
	\label{fig:apartIFC}
\end{figure}

Following the current IFC schema, to model apartment (or any cadastral unit), designers should use \textit{IfcBuilding}\footnote{\url{https://standards.buildingsmart.org/IFC/RELEASE/IFC4\_1/FINAL/HTML/schema/ifcproductextension/lexical/ifcbuilding.htm}}\textit{-CompositionType:`partial'} composed by \textit{IfcBuildingStorey}\textit{-CompositionType:} either \textit{`partial'}, \textit{`element'} or \textit{`complex'} (\Reffig{fig:apartIFC}).

The floor area and volumes are supposed to be extracted from there (by the BIM software itself or by different tools), either gross or net, and stored in the respective attributes `\textit{GrossPlannedArea}' and `\textit{NetPlannedArea}' in the \textit{BuildingCommon} property set associated to \textit{IfcBuilding}.
In addition, in the `\textit{BuildingUse}' property set, attributes are foreseen to store the intended functions of the building and of the apartment under the `\textit{MarketCategory}' label, for which, however, a strict enumeration is not foreseen.
Such a function is useful to check many regulations, among which the parking regulation treated in this project, but also others, such as the dimension regulation in other zones.
For example in the Rotterdam zone `\textit{Centre 2}'\footnote{\url{https://www.ruimtelijkeplannen.nl/documents/NL.IMRO.0599.BP1054Waterstad-va01/r\_NL.IMRO.0599.BP1054Waterstad-va01.html\#\_4\_Centrum-2}} the allowed dimensions are calculated differently based on the intended use of the building units.

In IFC4 the representation of a more complex articulation of spaces was introduced by means of \textit{IfcSpatialElement}\footnote{\url{https://standards.buildingsmart.org/IFC/RELEASE/IFC4\_1/FINAL/HTML/schema/ifcproductextension/lexical/ifcspatialelement.htm}}, as a superclass of \textit{IfcSpatialStructureElement}.
The \textit{IfcSpatialElement} subclasses \textit{IfcExternalSpatialStructureElement} and \textit{IfcZone} were added as well.
In particular, \textit{IfcZone} is interestingly defined to model for example energy zones or fire safety related ones: \textit{``A spatial zone is a non-hierarchical and potentially overlapping decomposition of the project under some functional consideration. A spatial zone might be used to represent a thermal zone, a construction zone, a lighting zone, a usable area zone. A spatial zone might have its independent placement and shape representation''}\footnote{\url{https://standards.buildingsmart.org/IFC/RELEASE/IFC4\_1/FINAL/HTML/schema/ifcproductextension/lexical/ifcspatialzone.htm}}.
However, it would be inaccurate to use it for apartments or other kind of building units, because those are usually enclosed by walls and tangible structures.

The modeling of spaces as found in the case study models (Section~\ref{sec:spaces})
could help calculations, since a unique geometry mapping the useful volumes, and corresponding areas can be directly queried instead of being extracted by a processing (e.g.\ to compute the non-explicitly-represented volume between the represented walls of an apartment).
However, it would be semantically inaccurate with respect to the IFC data model.

In any case, a codelist, likely an extension to the \textit{IfcSpaceType} enumeration, could help storing the function and/or level of articulation of each space, which condition the rule applications (parameters often depend on the spaces functions) and the consequent automatic processing.
In \ref{ann1}, some suggestions are reported as a useful reference for the development of such enumeration.

\section{Discussion}\label{sec:disc}

The adopted bottom-up approach allowed the insight into readiness of digital models as available from practice to automate the building permit checks.
The outlined issues are common to most of BIMs generated in practice and solving them is the necessary starting point for the development of powerful and scalable methodologies, guidelines and tools supporting digital building permit.

This study began slow, since it was necessary to start from small basic points and experiment with the available data in order to achieve the necessary knowledge for formulating a suitable methodology, later implemented in a tool.
However, it was not possible to skip such part, where the main value of this work lays: several basic issues, common to many processes and many regulations, had to be tackled and the problems had to be well-defined.
The adopted approach allowed the achievement of a feasible solution, working in practice.

The tool developed within the project, although still in a demonstrator phase, is built on solid theoretical, open standard-based and data-based bases.
This makes it effective and able to support the checks of the considered regulation.
The tool was tested with several IFC models, besides the ones belonging to the two provided BIMs within the project, to check replicability.

The addressed regulations were chosen for requiring common checks and information extraction from the models: the check of the maximum height of buildings is quite common to any city plan, as well as the maximum extension and dimensions of the buildings.
The other extracted information, such as the storeys' profiles, can be a useful base to check several parts of common urban planning rules (such as overlaps, covered surface, building overhangs, distance from other buildings and boundaries and so on).
The extracted geometries can also be suitably used to quickly update the city maps.
Many other regulations in Rotterdam, in Europe and beyond can have advantage of the tool performances.

In the case we considered, an Information Delivery Manual (IDM)\footnote{\url{https://technical.buildingsmart.org/standards/information-delivery-manual/} Accessed 4th November 2021}, which is the method provided by buildingSMART to define the necessary data requirements for specific data exchanges, and proper data requirements were not provided to designers. However, this represents a common situation in most of cases in most of countries.
Therefore, such an analysis and inspection was a necessary step to understand the level of data readiness, due both to the designers confidence with IFC and modeling in BIM and to the way IFC is implemented in software.
In fact, a recent study has tested the software support for IFC, in importing and exporting, demonstrating that several irregularities and errors can be produced due to differences in implementing IFC within tools \citep{noardo2021reference}.
Many of the issues encountered were actually due to mistakes and inaccuracies in modeling and exporting IFC, that could be more controlled by means of an Information Delivery Manual, but which are often still at a more general level of using IFC.
Part of the value of this paper therefore lies in the description of this misalignment as an initial essential step.
The guidelines that we propose are a relevant starting point to draft an IDM in future, that, however, will require more time.
In addition, the use of IDM itself is still being improved and defined by buildingSMART, leaving space for future work implementing the findings of this paper into specific buildingSMART standards and procedures (MVD, IDM, Information Delivery Specifications\footnote{\url{https://technical.buildingsmart.org/projects/information-delivery-specification-ids/} and \url{https://github.com/buildingSMART/IDS} Accessed on 5th November 2021} and so on).
In fact, a complex process is necessary, starting from stakeholders' needs and requirement definition, explication of regulations and necessary information, data features and modeling practice, until the implementation in a software solution \citep{idm2010,idm2012}.
Although being a process in progress within some municipalities, including Rotterdam, the results are still far from being reached, due to several concurring causes to be solved beforehand (e.g. difficulty in re-planning the processes within municipalities, need of getting municipality officers involved in the process, need of re-skilling of practitioners, define regulations unambiguously and likely in machine readable languages, improve interoperability of data, which requires both guidelines and agreed standard use and implementation of fully supporting tools).
For this reason, the guidelines provided in this study represent a valid provisional replacement of a complete IDM, likely coming in the future.

Besides the IDM, suitable Model View Definitions should be developed, \ie\ the (implemented) solution designed by buildingSMART for software to export part of the modelled BIM compliant to a specified part of IFC.
However, a current limitation of MVD mechanism is the difficulty in implementing custom views, useful for specific use cases, besides the example provided by buildingSMART.
Moreover, few authoring software tools are able to use custom MVDs, that makes it even more challenging.
For this reason, tools able to fix and further process the IFC models after export to make them better suitable for automated regulations checks would also be a valuable option for the future developments.

Moreover, pointing out the use of IFC in practice can be considered in the development of the standard itself.
For example, the mapping of the IFC schema to national languages translation, would foster a more aware adoption of IFC-based tools in current practice.

A secondary focus was the insertion of the building model within the city context, in order to address specifically the parts of the regulations which require such connection, by means of the integration with geoinformation.
However, the use of the IFC models as produced by practice presented more issues than expected, delaying the second (integration) step, which will be treated in future work more extensively.


\section{Conclusions}\label{sec:concl}

The paper presents the investigation developed to support the implementation of a tool and a methodology able to assist the municipalities in building permits regulation checks by means of digital solutions.
Starting points were the specific needs of the municipality and the involved data, including the regulations and digital standardized datasets (mainly IFC BIM) used directly as produced by architectural firms in practice.
The approach, developed within a project with the Rotterdam municipality (NL) was completely bottom-up, referring to the data and practice experience as explained by the municipality operators.

Using specific and concrete data from a case study data sets, which are though representative of many models, allowed pointing out their most frequent issues.
A minimum useful list of proper guidelines about the aspects to care during the data production phase were formulated based on them.
First of all, human care by the designer is necessary. Therefore, an improvement in training and education of the involved actors (practitioners and operators) is necessary.
In addition, a tool to validate the models (semantics and geometries), following clearer constraints, definitions and criteria would be useful.

The explication of the models' actual features was useful to guide the choices made when implementing the tool proposed in this study.
The demonstrator works with initial solutions and promises to be further improved and scaled for actually supporting the municipality in the building permit task on a wider scale.
Moreover, it would be very helpful to developers and inspectors whether the models' features were both compliant to a previously delivered IDM and also systematically stored within attached metadata.

The extension of the research supporting digital building permit likely includes many possibilities, with the important condition of inter-sectoral and multidisciplinary collaboration to guarantee the success and applicability of future developments.

\section*{ACKNOWLEDGEMENTS}\label{ACKNOWLEDGEMENTS}
This study was funded by the Municipality of Rotterdam (resp. Hasim Tezerdi and Rosen Manbodh) (December 2019 to July 2020), as a pilot within the `Rotterdam digital city' project.
This project has also received funding from the European Research Council (ERC) under the European Union’s Horizon2020 Research \& Innovation Programme (grant agreement no. 677312, Urban modeling in higher dimensions) and Marie Skłodowska-Curie grant agreement No. 707404 (Multisource Spatial data Integration for smart City Applications).


\bibliographystyle{elsarticle-num}
\bibliography{RotterdamAC_Resub_Revised_June21}

\appendix

\section{Space classifications and functions codelists}\label{ann1}

The space classification topic is not easily solved.
The enumeration \textit{IfcSpaceTypeEnum}\footnote{\url{https://standards.buildingsmart.org/IFC/RELEASE/IFC4\_1/FINAL/HTML/schema/ifcproductextension/lexical/ifcspacetypeenum.htm}} is foreseen by the IFC model (v.4) to fill the \textit{PredefinedType} attribute of \textit{IfcSpace}, but at the moment it is quite elementary. 
However, a high number of classifications of spaces and functions exist, that can be reused from other standards and shared data models.
For example, useful enumerations are provided by the Open Geospatial Consortium (OGC) CityGML\footnote{\url{http://www.citygml.org}} standard, relevant for the representation of the information related to the city, and mostly used within both 3D city models and in the experiments of integration of the BIMs with geoinformation, also for the building permit checking.
Other enumerations are provided by the OGC Landinfra\footnote{\url{http://www.citygml.org}} standard, the scope of which is to represent land and civil engineering infrastructure facilities.
The INSPIRE data model is part of the European Directive for spatial information in Europe\footnote{\url{https://inspire.ec.europa.eu}} aimed at the cross border representation of cities and land for shared environmental goals through Europe and it also prescribes useful classifications.
Moreover, other ontologies and vocabularies published within the semantic web framework report terms that can be reused for this study aims.

It is important to consider those existing classifications because, on the one hand, they are supposed to be comprehensive of many useful terms, correctly  and unambiguously defined.
In addition, having shared classifications would be an advantage for the integration and conversions necessary to further automate the use case.

The enumerations assessed as most relevant are here listed:

\begin{itemize}
	\item CityGML v.2.0 codelist for Room ``class'' (e.g.\ habitation, administration, catering, traffic, security\ldots);
	\item CityGML v.2.0 codelist for Room ``function and usage'' (e.g.\ living room, bedroom, kitchen, reception, office, canteen\ldots);
	\item CityGML v.2.0 codelist for BuildingInstallation ``function and usage'' (e.g.\ balcony, arcade, winter garden, column, stairs\ldots);
	\item CityGML v.2.0 codelist for AbstractBuilding ``class'' (e.g.\ habitation, business, sport, security, storage, traffic, church institution\ldots);
	\item CityGML v.2.0 codelist for AbstractBuilding ``function and usage'' (e.g.\ rubbish bunker, barn, stall, alpine cabin, court, tax office, concert building, museum, water mill, toilet, fort, chapel, green house\ldots);
	\item Landinfra ``CondominiumUseType'' codelist (e.g.\ residential, office, retail, garage, other);
	\item Landinfra ``BuildingPartType'' codelist (e.g.\ condominiumMainPart; condominiumAccessoryPart; jointAccessFacility; jointOtherFacility);
	\item INSPIRE data model for Buildings\footnote{\url{https://inspire.ec.europa.eu/id/document/tg/bu}} ``CurrentUseValue'' codelist (e.g.\ residential; individualResidence, agriculture, office, trade\ldots);
	\item INSPIRE data model for Buildings\footnote{\url{https://inspire.ec.europa.eu/id/document/tg/bu}} ``BuildingNatureValue'' codelist;
	\item in the Getty Vocabulary Art and Architecture Thesaurus\footnote{\url{https://www.getty.edu/research/tools/vocabularies/aat/about.html}}, as subclasses of ``single built works by functions''\footnote{\url{https://www.getty.edu/vow/AATHierarchy?find=church\&logic=AND\&note=\&subjectid=300004894}};
	\item others related to real estate fields\footnote{\url{https://www.archibus.net/ai/abizfiles/v21.2\_help/archibus\_help/Subsystems/webc/Content/web\_user/space/inventory/bldg\_perform/room\_category\_import\_standards.htm}}
	\item others related to facilities representation and management\footnote{\url{https://nces.ed.gov/pubs2006/ficm/content.asp?ContentType=Section\&chapter=4\&section=1\&skip=chapter\&skiptitle=4\%2E+Space+Use+Codes}}
\end{itemize}

Some of those enumerations include a very high number of values, intended to cover general and wide description aims.
Probably, not all of them are necessary to check the regulations or to serve design purposes.

The values used in the regulations and in the models themselves are here listed, in \Reftab{tab:parkingfunctions} and \Reftab{tab:BIMsfunctions} to be reference for future comprehensive works on this topic.

The need of a more complex study to propose a suitable classification also comes from the necessity of structuring it according to several criteria.
One of these is the hierarchy in functions (\eg\ under the category ``shop'' many kinds of shops can be included, such as small showrooms or commercial centers).
This is the case of the classification needed for the parking calculation, which is related to building units (apartment, shop, office, etc.).
This one can be in turn related to different levels of a meronymic (part-of) structure of spaces and aggregations possibly enclosed within spaces, as it happens in the two examples of BIMs for this project.
Therefore, it is possible to find, for example (from the small to the large): part of the room - the single room - the building unit - the building storey - possibly larger aggregations.
Third source of complexity, each of these parts could be defined for specific use cases, and they would have different configurations based on specific needs.
Some examples are the different zones for energy analysis, the volumes enclosing the walls for gross floor area calculation, the spaces including the spaces above the compound ceiling or not, which is possible to find now in the inspected BIM, and so on.

\begin{table*}
	\centering
	\small
	\begin{tabular}{|m{4cm}|m{10cm}|}
		\hline
		\textbf{Function} & \textbf{Sub-function} \\ \hline
		home	&	 \\ \hline
		to work	&	 \\ \hline
		&	office with counter function (banks, post offices) office 	 \\ \hline
		&	labor intensive / visitor intensive company (industry, laboratory, workshop etc.) 	 \\ \hline
		&	labor-intensive / visitor-extensive company (warehouse, storage, transport company, etc.) 	 \\ \hline
		&	multi-company building  \\ \hline
		shop	&	 \\ \hline
		&	store  \\ \hline
		&	convenience store	 \\ \hline
		&	large-scale retail trade	 \\ \hline
		&	hardware store / garden center / thrift shop	 \\ \hline
		&	showroom	 \\ \hline
		sports and recreation	&	 \\ \hline
		&	indoor gym / sports hall	 \\ \hline
		&	outdoor sports ground (per ha net)	 \\ \hline
		&	dance studio / gym	 \\ \hline
		&	squash court (per court)	 \\ \hline
		&	tennis court (per court)	 \\ \hline
		&	golf course (per hole)	 \\ \hline
		&	bowling center / billiard room (per court / table)	 \\ \hline
		&	swimming pool (per 100 m2 surface basin)	 \\ \hline
		&	allotments / utility gardens (per garden)	 \\ \hline
		&	riding school (per box)	 \\ \hline
		&	marina (per berth)	 \\ \hline
		&	event hall / exhibition building / conference building	 \\ \hline
		&	indoor playground / hall	 \\ \hline
		culture	&	 \\ \hline
		&	museum / library	 \\ \hline
		&	cinema / theater / theater (per seat)	 \\ \hline
		&	social cultural center / community center / funeral home religious building (church, mosque, etc.) per se. / visitor pl.)	 \\ \hline
		&	cemetery (by simultaneous funeral)	 \\ \hline
		catering industry	&	 \\ \hline
		&	cafe / bar	 \\ \hline
		&	restaurant	 \\ \hline
		&	hotel (per room)	 \\ \hline
		&	disco / party room	 \\ \hline
		&	cafeteria / snack bar	 \\ \hline
		Education	&	 \\ \hline
		&	cr\`eche / playgroup / nursery / after-school care 	 \\ \hline
		&		 \\ \hline
		&	primary education (per classroom of 30 ll)	 \\ \hline
		&	preparatory daytime education (vmbo, havo, vwo, per 30 ll classroom)	 \\ \hline
		&	vocational education day (MBO, ROC, HBO, WO) 	 \\ \hline
		healthcare 	&	 \\ \hline
		&	hospital (per bed) 	 \\ \hline
		&	nursing home (per residential unit) 	 \\ \hline
		&	pharmacy 	 \\ \hline
		&	1 e health care line (general practitioner, dentist, therapist) (per treatment room) 	 \\ \hline
	\end{tabular}
	\caption{Functions listed in the parking regulation.}%
	\label{tab:parkingfunctions}
\end{table*}

\begin{table*}
	\centering
	\small
	\begin{tabular}{|l|l|l|}
		\hline
		\textbf{Level of specification 1} & \textbf{Level of specification 2} & \textbf{Level of specification 3} \\ \hline
		area&	apartment / woonfunctie	&gang --- corridor	\\
		&	commerciele ruimte	&leftenhall --- lift hall	\\
		&	bijeenkomstruimte --- meeting room&	leidingschacht (toegankelijk)- pipe shaft (accessible)	\\
		&	parkeren&	toilet	\\
		&	stallingruimte (garage)	&brandweerlift --- fire lift	\\
		&	stallingruimpte (fietsen)	&liftkool --- elevator cabbage	\\
		&	diversen&	hoofdentree (tochtsluis) --- main entrance (draft lock)	\\
		&	&	flatrek	\\
		&	&	hoogspanningsruimte --- high voltage room	\\
		&	&	inkoopruimte --- purchasing space	\\
		&	&	laagspanningruimte --- low voltage space	\\
		&	&	containerruimte	\\
		&	&	in- en uitrit PG	\\
		&	&	pompruimte sprinkler	\\
		&	&	stadsverwarming --- district heating	\\
		&	&	sprinklerruimte	\\
		&	&	fietsenstalling	\\
		&	&	balcony	\\
		&	&	terras	\\
		&	&	woonkamer	\\
		&	&	slaapkamer	\\
		&	&	loggia	\\
		&	&	badkamer	\\
		&	&	berging	\\
		&	&	hal	\\
		&	&	garderobe	\\
		\hline
	\end{tabular}
	\caption{Labels assigned to \textit{IfcSpaces} in the two BIM models of the case study.}%
	\label{tab:BIMsfunctions}
\end{table*}

An investigation crossing a top-down approach consisting in the review and mapping of available classifications and a bottom-up approach considering the used values in regulations and models will be necessary to propose a working reference list of values.

\end{document}